\documentclass[12pt]{article}
\newcommand{\Letter}{
  \setlength{\textwidth}{7in}
   \setlength{\textheight}{9.5in}
  \hoffset=-0.75in
   \voffset=-1.15in }

\Letter

\begin{document}
%\addtolength{\baselineskip}{.5mm}
%\newlength{\extraspace}
%\setlength{\extraspace}{1.5mm}
%\setlength{\extraspaces}{2mm}

%\begin{document}

%\newcommand{\D}{\overline {\rm D}}
\def\D3{\overline{\rm D3}}
\begin{titlepage}
\begin{center}

{\hbox to\hsize{
\hfill COLO-HEP-536,~SU-ITP-08/01,~SLAC-PUB-13065}}

\vspace{3cm}

{\large \bf A Gravity Dual of Metastable\\[2mm]
Dynamical Supersymmetry Breaking}\\[1.5cm]

{Oliver DeWolfe${}^1$, Shamit Kachru${}^2$ and Michael Mulligan${}^2$}\\[8mm]

{${}^1$ \it Department of Physics, 390 UCB, University of Colorado, Boulder, CO 80309}\\[2mm]
{${}^2$ \it Department of Physics and SLAC, Stanford University, Stanford, CA
94305/94309}

\vspace*{1.5cm}

%%%%%%
{\bf Abstract}\\

\end{center}
\noindent
Metastable, supersymmetry-breaking configurations can be created in flux geometries by placing antibranes in warped throats.  Via gauge/gravity duality, such configurations should have an interpretation as supersymmetry-breaking states in the dual field theory. In this paper, we perturbatively determine the asymptotic supergravity solutions corresponding to $\D3$-brane probes placed at the tip of the cascading warped deformed conifold geometry, which is dual to an $SU(N+M) \times SU(N)$ gauge theory.  The backreaction of the antibranes has the effect of introducing imaginary anti-self-dual flux, squashing the compact part of the space and forcing the dilaton to run.  Using the generalization of holographic renormalization to cascading geometries, we determine the expectation values of operators in the dual field theory in terms of the asymptotic values of the supergravity fields.

%\vfill
%\hrulefill\hspace*{4in}

%{\footnotesize $^{\dagger}$ Email:
%\parbox[t]{7in}{}}

%\tableofcontents
%\newpage
\end{titlepage}
\newpage

\section{Introduction}

There are a number of reasons to be interested in theories which break supersymmetry at an exponentially low scale.  They may be of use as hidden sectors in supersymmetric models of weak-scale physics.  The dynamics involved in the supersymmetry breaking is also often intricate, exhibiting novel phenomena in field theory.  

One relatively new tool to study such theories is gauge/gravity duality \cite{Juan}.  Type IIB string theory on spaces of the form $AdS_{5} \times X^5$ with $X^5$ a five-dimensional Sasaki-Einstein space is known to be dual to a class of four-dimensional superconformal field theories  \cite{AdSreview}.
Simple examples with varying amounts of supersymmetry can be generated as orbifolds of $S^5$ \cite{orbifolds}, or starting from conical hypersurface singularities like the conifold \cite{KW}.  The dual field theories can be constructed by studying the gauge theories on $N$ D3-branes at the appropriate singularities.

Of more interest to us will be the backgrounds which are dual to ${\it confining}$ ${\cal N}=1$ gauge theories in four dimensions. A canonical example is provided by the theory of $M$ fractional D5-branes and $N$ D3-branes on the conifold, with $M \ll N$.  This theory exhibits a ``cascade" of Seiberg dualities \cite{Seiberg}, ending with chiral symmetry breaking at an exponentially small scale.
The dual supergravity solution, found by Klebanov and Strassler \cite{KS}, 
is completely smooth; the ``tip'' of the AdS-like throat is rounded off by
a complex deformation, whose magnitude is related to the scale of chiral symmetry breaking.

It was proposed in \cite{KPV} that one can make metastable supersymmetry breaking states starting from
such a geometry, by adding a small number $p \ll M$ of $\D3$-branes to the background.  In the probe
approximation, one can study the dynamics by using the DBI action on the antibranes.  They are attracted to the smooth tip of the geometry by the background 5-form flux and warping of the metric.  
The $\D3$-branes undergo the Myers effect \cite{Myers}, but for sufficiently small
$p/M$ they cannot perturbatively decay to a supersymmetric vacuum.  Instead,
the leading decay channel is via quantum bubble nucleation, which is highly suppressed for suitable choices of parameters.  Because the warp factor at the tip is exponentially small, these metastable states naturally break supersymmetry at an exponentially small scale, and thus should be usefully thought of as string duals of dynamical supersymmetry breaking.  These supersymmetry-breaking vacua are still largely mysterious from the field theory point of view.

For many purposes, it would be useful to have a supergravity solution which captures the long-range fields of the $\D3$-brane perturbation to the Klebanov-Strassler geometry.
Most notably, in gauge/gravity duality, one reads off the properties of states in a field theory by studying the (normalizable modes of) background supergravity fields.  So if one wishes to compute the vacuum energy and the VEVs of the operators in this supersymmetry-breaking state of the Klebanov-Strassler field theory, one should  find the supergravity solution in the presence of the $\D3$-branes and read off these quantities from the behavior of various fields at infinity.\footnote{A distinct class of non-supersymmetric vacua has been explored in \cite{KuSo}, building on the work of \cite{GuBo}.    
The theory investigated in \cite{KuSo} involves a perturbation of the Klebanov-Strassler field theory Lagrangian by a non-supersymmetric gaugino
mass term, which maps to a non-normalizable supergravity perturbation.  In contrast, we are studying states of the Klebanov-Strassler
theory itself, and only consider normalizable perturbations on the gravity side.   Other work exploring similar issues in closely
related theories appears in \cite{ABFK, DKS, Douglas, Dasgupta}.}

In this paper, we find the supergravity solutions characterizing these states, in perturbation theory
in $p/N$ and far from the tip of the throat.  More precisely, we linearize the IIB equations of motion around the large $r$ limit of the Klebanov-Strassler solution (first found by Klebanov and Tseytlin \cite{KT}), and identify the modes which are activated by the presence of the $p$ $\D3$-branes.  To make the computations more tractable, we study a ``smeared" solution, where we imagine the $\D3$-branes are spread uniformly over the internal dimensions at the tip.  

The result is simple to describe: in the supersymmetric theory, the geometry of the internal dimensions
is the 
Sasaki-Einstein space $T^{1,1}$, the background three-form flux $G_3$ is purely of type $(2, 1)$ relative to the complex structure, and the dilaton is constant.  We find
that the antibrane states are dual to solutions where the $T^{1,1}$ has been ``squashed" (in a way
that preserves the full global symmetry), the $G_3$ flux also acquires components of Hodge type $(1, 2)$, and the dilaton is forced to run.  Additionally, an energy density is generated in the field theory dual consistent with Poincar\'e invariance.
The squashing, the $(1, 2)$ flux, the dilaton and the energy density  are  all normalizable perturbations of the original theory; they describe a metastable state in the same supersymmetric field theory.
Using this solution and the holographic dictionary for the KS background determined by  \cite{ABY}, we are able to calculate the corresponding expectation values of the relevant operators in the SUSY-breaking state of the dual field theory.

The organization of the paper is as follows.  In \S2, we warm up by studying linearized supergravity equations
in a background of the form $AdS_5 \times X^5$.  While in a conformal theory one cannot find analogous metastable
states, many of the features of the solutions to the linearized equations will carry over to the confining case.
In \S3, we present and solve the linearized IIB equations for the background of interest, obtaining the
coefficients of the linearized perturbations by matching to antibrane sources, and describe the features of the solution as well as examining its contribution to the force on a probe D3-brane.
\S4 contains 
a discussion of the corresponding supersymmetry breaking in the dual field theory.
We conclude with a discussion
of directions for future exploration in \S5.
The appendices contain several calculations whose results we cite in the body of the paper.

\section{Perturbations to Conformal Backgrounds}
\label{ConformalSec}

Our ultimate interest 
is in $\D3$-brane-perturbed, supersymmetry-breaking configurations in the non-conformal background of the Klebanov-Strassler solution.
Before tackling those solutions, however, we warm up by considering an analogous computation in the simpler case of an AdS background.

\subsection{Anti-de Sitter backgrounds}

The background geometry around which we perturb is the $AdS_{5} \times X^{5}$ solution to type IIB string theory, where $X^5$ is an Sasaki-Einstein space.  (In fact any Einstein space will work, although a non-Sasaki-Einstein $X^5$ preserves no supersymmetry.)
The nontrivial fields are the metric and RR 5-form,
\begin{eqnarray}
\label{AdSSoln}
ds^2 = {r^{2} \over R^{2}} \eta_{\mu\nu} dx^\mu dx^\nu + {R^{2} \over r^{2}} (dr^2 + r^2 d\Omega^2_X) \,, \quad \quad \tilde{F}_5 = (1 + *) {4 r^{3} \over g_s R^{4}} dr \wedge {\rm vol}_{\rm R^4} \,,
\end{eqnarray}
where we defined ${\rm vol}_{\rm R^4} \equiv dx^0\wedge dx^1\wedge dx^2\wedge dx^3$, and $R^{4} \equiv 4 \pi^4 g_{s} N /{\rm vol}(X^5)$ with $N$ the quantized five-form flux, having set $\alpha'=1$.  We use the conventions of \cite{PS}.  The three-form field strengths $F_3$ and $H_3$ vanish, and the axio-dilaton field $\tau \equiv C_0 + i e^{-\Phi}$ is a 
modulus
that can take any constant value, with $e^{\Phi} = g_s$ the string coupling.  For $g_s$ small with 
fixed 't Hooft coupling $g_{s} N$,
string corrections are suppressed, and furthermore for $g_s N$ itself large we are in the supergravity regime with $\alpha'$ corrections negligible.

The field theory duals to these solutions are conformal field theories (CFTs), the isometry group $SO(4,2)$ of the $AdS_5$ space being realized as the four-dimensional conformal group.  In known examples the flux parameter $N$ is related to the rank of the dual gauge group(s).  For example, when $X^5 = S^5$ the dual is ${\cal N} = 4$ super-Yang Mills with $SU(N)$ gauge group, while for $X^5 = T^{1,1}$ the dual is an ${\cal N}=1$ supersymmetric $SU(N) \times SU(N)$ theory with bifundamentals and a superpotential.  We will review this example more extensively when we generalize it to the non-conformal case in section~\ref{CascadeSec}.

We are interested in perturbing these spacetimes with the introduction of space-filling $\D3$-branes.  These $\D3$-branes will fill the noncompact directions, and thus four-dimensional Poincar\'e symmetry is preserved.
A single brane would then sit at a point in both the radial direction $r$ and the angular directions on $X^5$; for simplicity, however, we will consider $p$ branes smeared over the compact space with a uniform density.  This preserves all the isometries of $X^5$ and hence provides a more elementary solution.

Since the total D3-brane charge associated to the 5-form is related to 
a parameter in the dual field theory, adding $\D3$-branes alone to a given background would result in 
a change to which field theory is the dual,
for example from $SU(N)$ to $SU(N-p)$.  In order to remain within a fixed theory as we perturb, we accompany the $p$ $\D3$-branes with $p$ D3-branes so that the total charge 
at infinity
is unchanged.  We place these branes at the same location as the antibranes, neglecting in the supergravity background the tachyon mediating brane/antibrane annihilation.

The AdS case has a number of features related to the conformal properties of the dual that actually make the solution to this ansatz more subtle to interpret than what we will find for the non-conformal case in section~\ref{CascadeSec}, despite being technically simpler.   The first of these is that we cannot
obtain static solutions with $\D3$-branes located at any nonzero value of the radial coordinate $r$
of the AdS spacetime, because the combination of the background warp factor and 5-form pulls them towards $r=0$; at best for $r > 0$ one could expect to find a time-dependent (non-Poincar\'e-invariant) solution, with the $\D3$-branes rolling towards $r=0$.  Thus we cannot include explicit brane sources that are static at a finite $r=r_0$, as it would lead to a non-conserved energy-momentum tensor and hence to inconsistent Einstein equations.  (This is in contrast to the non-conformal case, where 
static
$\D3$-branes at a nonzero $r_0$ will be possible.)

The alternative, having $\D3$-brane sources at $r=0$, is physically admissible but impossible to treat directly in perturbation theory in $1/r$.  Instead, we shall proceed by solving the homogeneous fluctuation equations without sources, focusing on the solution far from any branes, and then attempting to match to a $\D3$-brane solution by calculating the mass and charge at infinity.   A direct calculation of the mass will
work well for the non-conformal case; for the case at hand we find that the mass vanishes, and we will need to look at the limit of branes in flat space to justify the solution.  This limit will also provide evidence for the identification of the solution as indeed associated to antibranes at $r=0$, and will generate for us the higher-order corrections in $1/r$.

The other subtlety related to conformal invariance is that in a CFT, the vanishing of the trace of the stress tensor:
\begin{eqnarray}
\label{NoTrace}
T^\mu_\mu = 0 \,,
\end{eqnarray}
is an operator equation, and consequently applies to every state in the theory.
Any purported static, SUSY-breaking
antibrane state, on the other hand, preserves Poincar\'e invariance and so any VEV for the stress tensor must obey $\langle T_{\mu\nu} \rangle = t \eta_{\mu\nu}$ for some $t$.  But then the tracelessness condition (\ref{NoTrace}) requires $t=0$; there is no way to preserve Poincar\'e invariance in a CFT while generating an energy density. (Of course, dual geometries like a black brane may source the stress tensor; these geometries have a horizon and so put time on a different footing, thereby breaking the Poincar\'e symmetry and avoiding contradiction with this discussion.)

This would be a surprise if we expected a state dual to static $\D3$-branes at a nonzero radial coordinate $r_0$, since one expects such a state to break SUSY at a finite energy scale corresponding to the radial position, and in a globally supersymmetric theory such a breaking must lead to an energy density.  As we just argued, however, solutions can only correspond to static antibranes if they are all the way down the throat at $r=0$, corresponding to a vanishing energy scale.  We shall indeed see that the only solution consistent with our ansatz for the AdS case leads to no energy density in the dual, and we will give an interpretation of the solutions as corresponding to
brane/antibrane pairs at $r=0$.

\subsection{Perturbation to the anti-de Sitter background}

Perturbing about this background, we consider an ansatz that preserves the full $SO(3,1)$ Poincar\'e symmetry.
Moreover, the branes are smeared over the compact space, thereby preserving
the isometries of $X^5$.  For the specific case of $X^5 = S^5$ this preserves a full $SO(6)$ and requires that the metric and 5-form ansatz take the form,
\begin{eqnarray}
\label{N4Ansatz}
ds^2 = e^{2A(r)} \eta_{\mu\nu} dx^\mu dx^\nu + e^{2B(r)} dr^2 + e^{2B(r)} r^2 d\Omega^2_X \,, \quad \tilde{F}_5 = (1 + *) \, d{\alpha(r) \over g_s} \wedge dx^0\wedge dx^1\wedge dx^2\wedge dx^3 \,,
\end{eqnarray}
where $d \Omega_X^2$ is the Einstein metric on $X^5$.

For other $X^5$ the isometry group is smaller and in principle a more general ansatz could be considered, but we shall keep this ansatz as the only one that applies to all AdS backgrounds, expecting that the smeared antibrane solution will be universal across all choices of $X^5$.  (We will make use of an ansatz that has more freedom to both warp and squash the compact space in section~\ref{CascadeSec}.)

We also take the 3-form field strengths $F_3$ and $H_3$ to vanish, as they would violate the global symmetry  in the $S^5$ example, and in any case are not sourced directly by the $\D3$-branes.  This in turn requires the axion/dilaton, another field not sourced directly by the antibranes, to be harmonic according to its field equation, and we choose it to remain constant.  (A decoupled fluctuation of the dilaton alone also exists, but we will not write it down; a similar mode will show up in the non-conformal case.)
In principle we could have had different coefficients $e^{2B}$, $e^{2C}$ for the $dr^2$ and $d\Omega_X^2$ terms, but we used our coordinate freedom to redefine $r$ so as to set $B=C$, which will simplify the calculations.

We turn now to the field equations.
Consider first the 5-form.  Its Bianchi identity (equivalent to the equation of motion by self-duality) gives
\begin{eqnarray}
\partial_r ( e^{-4A+4B} r^5 \partial_r \alpha) = \sum_{{\rm 3-branes}} \pm r^5 (2 \kappa_{10}^2 T_3) {\delta^6(y) \over \sqrt{\tilde{g}}} \,,
\end{eqnarray}
where the sign of the right-hand-side depends on whether a brane is D3 or $\D3$, $y$ runs over
the six  transverse dimensions to the Minkowski slices, and $\tilde g$ is the unwarped metric on the transverse six-space
$ds^2 = dr^2 + r^2 d\Omega_{X}^2$.  We can integrate both sides:
\begin{eqnarray}
\int d\Omega \int dr \partial_r ( e^{-4A+4B} r^5 \partial_r \alpha) = (2 \kappa_{10}^2 T_3) N_{enc} \,,
\end{eqnarray}
where $N_{enc}$ is the charge enclosed out to radius $r$, and so
\begin{eqnarray}
\label{FiveFinal}
 \partial_r \alpha = r^{-5} e^{4A-4B} {2 \kappa_{10}^2 T_3 N_{enc} \over {\rm vol}(X)} \,.
\end{eqnarray}
Using $2 \kappa_{10}^2 T_3 = g_s (2 \pi)^4$ we verify that this is satisfied for an unperturbed AdS solution.

Meanwhile, we obtain the following independent 
Einstein equations.  First from the non-compact components,\begin{eqnarray}
\label{MinkEqn}
A'' + 4 {A'}^2 + 4 A'B' + {5 \over r} A' =  {1 \over 4}  e^{-8A}  (\partial_r \alpha)^2 + {1 \over 2} e^{-4B} (\kappa_{10}^2 T_3) \sum_{3-{\rm branes}} {\delta^6(y) \over \sqrt{\tilde{g}_6}} \,,
\end{eqnarray}
while from the $rr$ components,
\begin{eqnarray}
\label{RadialEqn}
 -4A'' - 5 B'' -4 {A'}^2  +4 A' B'  - {5 \over r} B' = - {1 \over 4} e^{-8A} (\partial_r \alpha)^2  +  {1 \over 2}e^{-4B} (\kappa_{10}^2 T_3) \sum_{3-{\rm branes}} {\delta^6(y) \over \sqrt{\tilde{g}_6}} \,,
\end{eqnarray}
and finally from the compact components,
\begin{eqnarray}
\label{AziEqn}
B'' + 4{B'}^2 + 4 A' B'  + {4 \over r} A'+ {9 \over r} B' = -{1 \over 4} \, e^{-8A} (\partial_r \alpha)^2  -{1 \over 2} e^{-4B} (\kappa_{10}^2 T_3) \sum_{3-{\rm branes}} {\delta^6(y) \over \sqrt{\tilde{g}_6}} \ .
\end{eqnarray}
In the Einstein equations
both D3-branes and $\D3$-branes contribute with the same sign; we have kept the brane sources for completeness, but will ultimately drop them.  It is also useful to take a linear combination of these equations to remove the $A''$ and $B''$ terms; the combination $4 (\ref{MinkEqn}) + (\ref{RadialEqn}) + 5(\ref{AziEqn})$ accomplishes this, and happens to remove the $\delta$-function sources as well.  The result is the relation
\begin{eqnarray}
\label{Constraint}
12 {A'}^2 + 20 {B'}^2 + 40 A' B' + {40 \over r} A' + {40 \over r} B' = - {1 \over 2} e^{-8A} (\partial_r \alpha)^2 \,.
\end{eqnarray}
Consider the linearization of the above equations to first order in fluctuations about the background AdS solution (\ref{AdSSoln}) $e^{4A_0} =  e^{-4B_0} = \alpha_0 = r^4/R^4$. We write
\begin{eqnarray}
A = \log {r \over R} + \delta A \,, \quad
B = -\log  {r \over R} + \delta B \,, \quad
\alpha = {r^4 \over R^4} + \delta\alpha \,.
\end{eqnarray}  
Examine first the integrated five-form equation (\ref{FiveFinal}).  The zeroth order part is satisfied.  At first order, we obtain
\begin{eqnarray}
\label{AlphaPertSoln}
\partial_r \delta \alpha = 16 {r^3 \over R^4} ( \delta A - \delta B) \,.
\end{eqnarray}
We shall use this to eliminate the 5-form contributions from the Einstein equations.  From now on we drop the delta-function sources, anticipating a solution far from the branes.  Linearizing the Einstein equations, we find:
\begin{equation}
\label{LinearMink}
\delta A'' + 9 {\delta A'\over r}   + 4{\delta B' \over r} + 32{ \delta B \over r^2} = 0 \,,
\end{equation}
\begin{equation}
\label{LinearRad}
4\delta A'' + 12 {\delta A' \over r}  + 5 \delta B'' + {\delta B' \over r} + 32 {\delta B  \over r^2} = 0 \,,
\end{equation}
\begin{equation}
\label{LinearAz}
\delta B'' + 5 {\delta B' \over r} -32 {\delta B \over r^2} = 0 \,,
\end{equation}
Also useful is the linearized version of the relation (\ref{Constraint}), which is a linear combination of these three,
\begin{eqnarray}
\label{LinearCon}
r \delta A'  + {5 \over 3} r \delta B' - {8 \over 3} \delta B= 0\,.
\end{eqnarray}
Of the three linearized Einstein equations (\ref{LinearMink}), (\ref{LinearRad}) and (\ref{LinearAz}) only two are independent thanks to the coordinate invariance of the theory.  We choose as our two independent equations the Einstein equation with indices on the compact space (\ref{LinearAz}) as well as the relation (\ref{LinearCon}); it is easy to check they imply the other two equations (\ref{LinearMink}) and  (\ref{LinearRad}).  

Equation (\ref{LinearAz}) is a linear second-order equation for $\delta B$ only, and has the obvious power-law solutions
\begin{eqnarray}
\label{AdSFluct1}
\delta B = B_{4} r^4 + B_{-8} r^{-8} \,.
\end{eqnarray}
The relation (\ref{LinearCon}) then determines the mode $\delta A$ up to a constant:
\begin{eqnarray}
\label{AdSFluct2}
\delta A= A_0 - B_4 r^4 - 2 B_{-8} r^{-8} \,,
\end{eqnarray}
and any $A_0$ may be absorbed into a redefinition of the 4D coordinates.  Thus  (\ref{AdSFluct1}), (\ref{AdSFluct2}) with the 5-form determined by (\ref{AlphaPertSoln}) constitute the unique solution to the fluctuation equations around an AdS background consistent with the symmetries we have imposed.

The scalings of $r^4$ and $r^{-8}$ for this mode are consistent with a fall-off we expect for the non-normalizable and normalizable solutions of a field associated to a $\Delta = 8$ operator ${\cal O}_8$ in the field theory dual.  Such a mode is known to exist in the KK spectrum of $AdS_5 \times S^5$ as well as $AdS_5 \times T^{1,1}$; indeed it is a universal fluctuation present for all $AdS_5 \times X^5$ geometries \cite{Stability}, owing to the fact that it is constant over the compact space and hence is indifferent to the details of the spectrum of harmonics.
In the ${\cal N}=4$ case, this operator is known to have the form
\begin{eqnarray}
{\cal O}_8 = {\rm Tr} \ ( W_\alpha^2 \bar{W}_{\dot\beta}^2 ) + \ldots \sim {\rm Tr}\ (F_{\mu\nu})^4 + \ldots
\end{eqnarray}
Thus the leading fluctuation in the metric consistent with the symmetries corresponds to this universal $\Delta = 8$ operator ${\cal O}_8$.  The $r^4$ mode 
is non-normalizable,
blowing up as the boundary is approached,
and corresponds to 
a perturbation of the field theory dual by the ${\cal O}_{8}$ operator.
Hence the leading normalizable perturbation to the background $AdS_{5} \times S^{5}$ geometry is the $r^{-8}$ mode:
\begin{eqnarray}
\label{Delta8}
\delta B = B_0 r^{-8}  \,, \quad \quad \delta A = - 2 B_0 r^{-8} \,.
\end{eqnarray}
This 
corresponds to
an expectation value for the operator ${\cal O}_8$.

Let us note an operator that is {\em not} sourced or given a VEV by this solution: the stress tensor $T_{\mu\nu}$.  As already described, there is no nonzero value of $T_{\mu\nu}$ consistent with both Poincar\'e symmetry and conformal invariance, and hence finding zero is what we would expect.  Thus there is a vanishing energy density in the field theory dual, and hence unbroken supersymmetry, for any value of the
coefficient of the linearized solution.    

The lack of energy density can also be seen by considering how this perturbation affects the appropriate generalization of the ADM mass for anti-de Sitter space.  Using the formula from Appendix A, we discover that this mass vanishes:
\begin{eqnarray}
\label{AdSADM}
E_{ADM} \propto \lim_{r \to \infty} B_0 r^{-4} \; \to \;  0 \,,
\end{eqnarray}
and so this perturbation does not change the total mass of the geometry as compared with the background AdS space.  In general only a $\Delta = 4$ operator has the correct scaling to contribute a finite amount to the ADM mass.

Since supersymmetry is not broken and the ADM mass is not shifted, 
connecting solutions with $\langle {\cal O}_8 \rangle \neq 0$ 
to brane/antibrane perturbations may seem like an obscure thing to do.  While such an interpretation
will be much clearer in the case of confining throats discussed next, it is however still possible
to formally derive our linearized solution from brane/antibrane probe perturbations, as follows.

Solutions for Poincar\'e-invariant, non-BPS $p$-branes asymptoting to flat space instead of anti-de Sitter space were obtained by Brax, Mandal and Oz \cite{BMO}.  In Appendix B, we demonstrate that the ${\cal O}_8$-perturbed solution found above can also be obtained as the near-horizon geometry of a configuration corresponding to $N+p$ D3-branes and $p$ $\D3$-branes all sitting at the origin, in a particular scaling limit; the coefficient $B_0$ is found to go as $B_0 \sim p/N$, as expected. 
The overall ADM mass, nonzero for the flat space solutions, is scaled to zero in this limit.  
This derivation also produces corrections to the perturbation found here subleading in $p/N$ and at higher order in $1/r$.

Thus it appears that the ${\cal O}_8$ solution is indeed related to brane/antibrane configurations, but ones with the perturbing branes sitting at $r=0$.  We may interpret the motion of the $\D3$-branes all the way down the throat as the gravity dual of brane/antibrane annihilation, since the energy scale in the dual has retreated to zero, and hence the supersymmetry breaking has disappeared.
This issue does not occur in the non-conformal case, to which we now turn.

\section{Antibranes in the cascading geometry}
\label{CascadeSec}

We now consider our primary interest --- studying supersymmetry-breaking antibrane perturbations of the cascading Klebanov-Strassler geometry associated to fractional branes on the conifold.  This background differs in two ways from the anti-de Sitter backgrounds.  First, conformal invariance is broken at all scales (the gravity solution differs from AdS at all values of $r$), and consequently there is nothing forbidding a Poincar\'e-invariant perturbation that generates a nonzero energy density.  And secondly, the dual theory is confining with a mass gap, realized in the gravity configuration by a smooth termination of the throat at a certain minimal warp factor at $r = r_0$. As a result, a $\D3$-brane may achieve a metastable static configuration by relaxing to $r_0$, corresponding to SUSY breaking at a nonzero but exponentially small scale in the field theory dual.  We thus expect that asymptotic gravity solutions 
which are Poincar\'e invariant and carry non-zero energy density (determined by the scale
of SUSY breaking) will be achievable, and indeed that is what we find.

In this section we obtain the linearized perturbation to the background associated with the addition of brane/antibrane pairs (arguing in \S3.2\ that this allows us to read off also the asymptotic solutions relevant
to the antibrane-only states of \cite{KPV}), and identify the nonzero energy density resulting from the supersymmetry breaking.  We discuss the geometry of the solution and the associated three-form flux, 
finding that the SUSY breaking results in a ``squashing" of the compact Einstein space $T^{1,1}$, a running of the dilaton and the generation of a
$(1, 2)$ component of the three-form flux.
We also discuss the fate of this perturbation in the conformal limit, finding consistency with
the results of \S2.

\subsection{The Klebanov-Tseytlin geometry and dual field theory}

The full geometry of the cascade solution is that of Klebanov and Strassler (KS) \cite{KS}, which consists of a warped deformed conifold; the deformation brings the geometry to a smooth end in the infrared, which is associated with confinement and a mass gap in the dual field theory.  

We shall focus on solutions at the UV end of the geometry, far from the brane sources which we place in their metastable configuration at the bottom of the throat. The UV end of the KS throat reduces to the solution of Klebanov and Tseytlin (KT) \cite{KT}.  Although the KT geometry does not include the smooth ending at the tip of the throat, instead continuing down to a singularity at small $r$, it remains non-conformal and must be able to support solutions corresponding to perturbations of the true KS throat.   Since the full KS geometry can smoothly bring $\D3$-branes to a halt at a
finite
radial distance corresponding to a nonzero energy scale in the dual, it is natural to expect that solutions in the UV exist corresponding to the large $r$ limit of the supergravity fields 
produced by
the antibrane 
configurations;
this is indeed with what we find. We will identify the perturbation by calculating the associated generalization of an ADM mass. 

The KT background is a deformation of the $AdS_5 \times T^{1,1}$ geometry associated to $N$ D3-branes on the conifold by three-form fluxes corresponding to the addition of $M$ fractional D5-branes. The metric is
 \begin{equation}
 \label{eq:KT}
 ds^2_{KT} = e^{2A(r)}\eta_{\mu \nu} dx^{\mu}dx^{\nu} + e^{-2A(r)}(dr^2 +r^2 e_\psi^2 + r^2 \sum_{i=1}^2(e_{\theta_i}^2 + e_{\phi_i}^2) ) \,,
 \end{equation}
where the one-forms associated to the compact space $T^{1,1}$ can be taken to be \,,
\begin{eqnarray}
e_\psi = {1 \over 3} (d \psi + \sum_{i=1}^2 \cos \theta_i d\phi_i ) \,, \quad e_{\theta_i} = {1 \over \sqrt{6}} d \theta_i \,, \quad e_{\phi_i} = {1 \over \sqrt{6}} \sin \theta_i d \phi_i \,,
\end{eqnarray}
for $i=1, 2$, and the warp factor is
\begin{eqnarray}
\label{eqn:unpertmetric}
 e^{-4A}  = {27 \pi g_s \over 4 r^4} \left(N +{ 3 g_s M^2 \over 2 \pi} (\log (r/r_{UV}) + {1 \over 4}) \right) \,,
\end{eqnarray}
where $r_{UV}$ is the scale at which the theory is defined.  
There is a naked singularity where 
$e^{-4A} = 0$,
which is simply a relic of the fact that the solution is only valid for sufficiently large distances from the tip of the cone; the complete KS solution removes this singularity via deformation.  
The self-dual 5-form flux is
\begin{eqnarray}
\tilde{F}_{5} = (1+\ast) {\cal F}_5, \quad \quad{\cal F}_5 = -27 \pi N_{eff}(r) {\rm vol}_{T^{1,1}} \,,  \quad N_{eff}(r) \equiv \left(N +{ 3 g_s M^2 \over 2 \pi} \log (r/r_{UV})  \right) \,,
\end{eqnarray}
where the volume form on $T^{1,1}$ is ${\rm vol}_{T^{1,1}} = e_\psi \wedge e_{\theta_1} \wedge e_{\phi_1} \wedge e_{\theta_2} \wedge e_{\phi_2}$.  Finally there is 3-form flux,
\begin{eqnarray}
F_3 = {9 M \over 2} e_\psi \wedge (e_{\theta_1} \wedge e_{\phi_1} - e_{\theta_2} \wedge e_{\phi_2}) \,, \quad \quad B_2 = {9 g_s M \over 2} (e_{\theta_1} \wedge e_{\phi_1} - e_{\theta_2} \wedge e_{\phi_2}) \log(r/r_{UV}) \,.
\end{eqnarray}
The axio-dilaton $\tau = C_0 + i e^{-\Phi}$ remains constant.

The preserved symmetries of this background are $SO(3,1)$ of the four-dimensional $x^\mu$ along with the $SU(2)_1 \times SU(2)_2 \times Z_{2M} \times Z_2$ of the $T^{1,1}$; the $SU(2)$s act on the two-spheres associated to $\theta_1$, $\phi_1$ and $\theta_2$, $\phi_2$, and the $Z_2$ exchanges them. The metric possesses a $U(1)_\psi$ acting on the $\psi$-coordinate, which is respected by the other field strengths but broken to $Z_{2M}$ by any concrete expression for the potential $C_2$ 
\cite{Klebanov-Ouyang-Witten}.
This is broken further to $Z_{2}$ in the full KS solution, reflecting the fermion condensate in the IR.

In the $M \to 0$ limit we revert to $AdS_5 \times T^{1,1}$, corresponding to $N$ D3-branes sitting at the tip of the conifold.  The dual field theory to this conformal limit is an ${\cal N} =1$ superconformal field theory, which can be realized as an $SU(N) \times SU(N)$ ${\cal N}=1$ gauge theory with bifundamental matter $A_i$ and $B_j$ in the $(N, \overline{N})$ and $(\overline{N}, N)$ of the gauge group.  $SU(2)_1 \times SU(2)_2$ is a global symmetry with $A_i$ and $B_j$ transforming as doublets of the first and second factor respectively, and moreover there is a superpotential
\begin{eqnarray}
\label{Superpotential}
W = \lambda \epsilon^{ik} \epsilon^{jl} \ {\rm Tr} \ A_i B_j A_k B_l  \,.
\end{eqnarray}
$U(1)_\psi$ is restored in the conformal limit as the R-symmetry. 

This superconformal theory has two complex moduli
parameterizing the space of marginal deformations,
which may be thought of as gauge couplings and theta angles for each $SU(N)$ factor; for given values of these couplings, the superpotential parameter $\lambda$ is adjusted to reach the fixed point, where $A_i$ and $B_j$ have dimension $3/4$ and the superpotential is marginal.  The complex moduli are dual on the gravity side to the axio-dilaton and to $\int_{S^2} (B_2 + i C_2)$, respectively,
and their associated operators, along with the energy-momentum tensor $T_{\mu\nu}$ dual to the four-dimensional metric, make up the exactly marginal ($\Delta = 4$) operators of the theory.

For nonzero $M$, the theory becomes an $SU(N) \times SU(N + M)$ gauge theory, with $A_i$ and $B_j$ still bifundamentals of the new gauge groups and with the same superpotential.  Conformal symmetry is broken, the R-symmetry $U(1)_\psi$ is broken to $Z_{2M}$ by anomalies, and one of the two moduli is lifted.  The simple way to think of this is that the sum of the gauge couplings $1/g_1^2 + 1/g_2^2$ remains a modulus dual to the constant value of the dilaton $\tau$, while the difference begins to flow with energy scale,
\begin{eqnarray}
\label{runaway}
{ 1\over g_{1}^{2}} - {1 \over g_{2}^{2}} \sim {1 \over g_s} \int_{S^{2}} B_{2} \sim M \, {\rm log}(r/r_{UV}) \,.
\end{eqnarray}
We shall refer to the operator associated to the sum of the gauge couplings as ${\cal O}_+$, and the one associated with the difference as ${\cal O}_-$.  As we discuss later, properly both couplings also involve the superpotential parameter $\lambda$.

The RG evolution exhibits the beautiful phenomenon of self-similar flow called the duality cascade, where the effective number of colors $N_{eff}$ diminishes as the scale decreases, corresponding to successive Seiberg dualities that progressively reduce the ranks of the gauge group $SU(N) \times SU(N + M) \to SU(N-M) \times SU(N)$ and so on, while preserving the structure of the theory at each step.

\subsection{Relation to metastable antibrane states}

Before proceeding, we should note the relation of these states to the states of  \cite{KPV}.  There, one considered a warped throat  to be part of a compact geometry, with  
\begin{equation}
\int_A F_3 = M,~~\int_B H_3 = -K  \,.
\end{equation}
Adding $p \ll M$ $\D3$-branes to this geometry was then argued to result in metastable supersymmetry-breaking vacua.  How does this compare to the case we study with $p$ ordinary D3-branes as well as the $p$ $\D3$-branes?

Consider the issue of total charges.  Because of the Bianchi identity for $\tilde{F}_5$, these fluxes contribute $MK$ units to the total D3-brane charge $Q_{\rm D3}$, which schematically reads:
\begin{equation}
\label{D3Charge}
Q_{\rm D3} = \int d \tilde  F_5 = N_{\rm D3} - N_{\D3} +  \int H_3 \wedge F_3 = N_{\rm D3} - N_{\D3} + MK \,,
\end{equation}
The tunneling decay mechanism of the $p$ $\D3$-branes must conserve this total charge.  This proceeds by the dynamical nucleation of an NS5-brane bubble which wraps the A-cycle; this brane acts as a source for $H_3$ and shifts $K \to K-1$.  This alone changes the total charge (\ref{D3Charge}) by $-M$, but the process compensates  by also creating $M$ D3-branes ending on the bubble (and filling the space ``outside").  $p$ of these newly-created D3-branes annihilate the antibranes, leaving a supersymmetric configuration with $M-p$ D3-branes, background flux giving rise to charge $N-M$, and no $\D3$-branes.  These together have D3 charge $N-p$, while the $F_3$ flux remains $M$.
Consequently, the $p$-$\D3$-brane state should be viewed as a metastable state, not in an $SU(N+M) \times SU(N)$ theory, but in the $SU(N+M-p) \times SU(N-p)$ supersymmetric gauge theory associated to $N-p$ D3s and $M$ fractional D5s at the conifold.

Imagine now a noncompact case where we let $N+ p$ be an integer multiple of $M$, instead of $N = KM$.
The asymptotic geometry we use to study this theory, after the addition of $p$ probe D3s and $p$ probe $\D3$s to the background characterized by $N$ (but forbidding the brane/antibrane tachyons to condense),  should have ${\it identical}$ conserved
charges compared to the theory of $N+p$ D3s and $M$ fractional 5-branes with $p$ $\D3$-brane probes.  
Since now $N+p$ is divisible by $M$, we are back in the situation considered in \cite{KPV}, and the asymptotic solution we find should agree with that for $p$ probe $\D3$-branes in the confining Klebanov-Strassler geometry as far as the asymptotic charges are concerned.  Therefore, if we find a solution with the correct values of the asymptotic charges, motivated by adding $p$ D3/$\D3$ pairs to the geometry with $N$ D3s and $M$ fractional D5s, we can be confident it coincides with the  supergravity solution for the metastable $\D3$-brane states.  

The reader may have found the preceding argument somewhat confusing; let us give a slightly different explanation
of the same fact.  We could formally find a supersymmetric solution with zero vacuum energy, by adding $p$ ``imaginary branes"
with negative D3 charge and negative D3 tension to the geometry with $N+p$ branes; it would be supersymmetric
because these ``imaginary branes'' satisfy a BPS condition relating their charge and tension equivalent to the relation satisfied by the $N+p$ ``real'' branes.  The solution we are interested
in differs from this one, by adding $p$ D3-branes and $p$ $\D3$-branes.  (The $p$ D3s can then cancel against the $p$ ``imaginary branes" as sources for charge and energy density).  Hence, our solution of interest should have an energy density characteristic of 
$2p$ (warped) D3s, and the same charge as the solution with $N$ units of D3 charge.  This argument was first given by Maldacena and Nastase in a rather different context (supersymmetry breaking in the gravity dual of a 3d ${\cal N}=1$ supersymmetric
Chern-Simons theory) \cite{Maldacena-Nastase}.

\subsection{Perturbation ansatz in KT background}

We now wish to consider the response of the cascading geometry when it is perturbed by a D3-brane/$\D3$-brane source.

One way to approach this problem would be to explicitly place branes and antibranes at the base of the full KS geometry and discover the response of the system in the UV.  This would entail, however, somewhat tedious matching between the IR and UV regions.  Instead we shall pursue the same philosophy as in the AdS case, and consider sourceless fluctuations in the UV regions --- that is, in the KT geometry --- and then match these to an ADM mass to verify that they correspond to the addition of tension without charge.  We shall also demonstrate explicitly that branes will source the particular fields that are turned on.

As in the conformal case of  the previous section, we will simplify matters by smearing the perturbing branes over the compact space such that the full symmetry of $T^{1,1}$ is maintained.  We thus preserve Poincar\'e invariance as well as $SU(2)_1 \times SU(2)_2 \times Z_2$, and we keep the metric $U(1)_\psi$ symmetric as well.  Our ansatz for the metric is thus constrained to be
\begin{equation}
\label{eqn:metric}
ds^{2} = r^{2}e^{2a(r)} \eta_{\mu \nu} dx^{\mu}dx^{\nu} + e^{-2a(r)} \left( {dr^{2} \over r^{2}} + e^{2b(r)} e_\psi^2 + e^{ 2c(r)} \sum_{i=1}^2(e_{\theta_i}^2 + e_{\phi_i}^2) \right) \,,
\end{equation}
where we have found it convenient to extract a factor of $r^2$ from the warp factor.  Additionally, we have used our freedom to perform coordinate redefinitions of $r$ to relate the coefficient of the $dr^2$ term with the coefficient of the Minkowski part.  

Notice that even maintaining all symmetries due to the smeared branes, it is possible to introduce a new metric function, the ratio $e^{b-c}$ between the length scale of the circle acted on by $U(1)_\psi$ and that of the two $S^2$s.  We shall find this degree of freedom to be essential in what follows.

For notational convenience we define the following rescaled constants:
\begin{eqnarray}
\bar{N} = 27 \pi N \,, \quad \bar{M} = {9 \over 2} M \,,
\end{eqnarray}
and for the remaining fields we take,
\begin{eqnarray}
F_{3} & = & \bar{M} e_\psi \wedge (e_{\theta_1} \wedge e_{\phi_1} - e_{\theta_2} \wedge e_{\phi_2}) \,, \quad
B_{2} =  k(r)(e_{\theta_1} \wedge e_{\phi_1} - e_{\theta_2} \wedge e_{\phi_2}) \,, \\
\tilde{F}_5 &=& d C_4 - C_2 \wedge H_3 \,, \quad \quad C_4 = {\alpha(r) \over g_s} \, {\rm vol}_{R^4} \,, \\
\Phi & = & \Phi(r), \quad \quad C_0=0 \,.
\end{eqnarray}
where again ${\rm vol}_{R^4} \equiv dx^0 \wedge dx^1 \wedge dx^2\wedge dx^3$.
For the five-form, we use the forms of the potentials and the Bianchi identity to obtain,
\begin{eqnarray}
\tilde{F}_5 = (1 + \ast) \, {\cal F}_5 \,, \quad \quad {\cal F}_5 =  - C_2 \wedge H_3 = - \bar{N}_{eff}(r) \, {\rm vol}_{T^{1,1}} \,, \quad \bar{N}_{eff}(r) \equiv \bar{N} + 2 \bar{M} k(r) \,,
\end{eqnarray}
and the potential $C_4$ is determined by self-duality to be
\begin{eqnarray}
\partial_r \alpha  =- g_s N_{eff}(r) \, r^3 e^{8a-b-4c} \,.
\end{eqnarray}
We now look for solutions to this ansatz.  The field equations of Type IIB SUGRA are \cite{PS},
\begin{eqnarray}
\label{OrigDilatonEqn}
\nabla^{2} \Phi & = & e^{2\Phi}\partial_{M}C_0 \partial^{M}C_0 - {g_s e^{-\Phi} \over 12}( H_{MNP}H^{MNP} - e^{2\Phi} F_{MNP}F^{MNP}) \\
\nabla^{M}(e^{2\Phi} \partial_{M}C_0) & = & - {g_s e^\Phi \over 6} H_{MNP} \tilde{F}^{MNP} \,, \quad d * \tilde{F}_{5}  =  - F_{3} \wedge H_{3}  \,,\\
d \ast (e^{\Phi} \tilde{F}_{3}) & = & g_ s \tilde{F}_{5} \wedge H_{3}\,, \quad
d \ast (e^{- \Phi} H_{3} - C_0 e^{\Phi} \tilde{F}_{3})  = - g_s \tilde{F}_{5} \wedge F_{3}  \,, \\
\nonumber
R_{MN} & = & {1\over 2} \partial_{M} \Phi \partial_{N} \Phi + {e^{2\Phi} \over 2}\partial_{M}C_0 \partial^{M}C_0 + {g^2 \over 96} \tilde{F}_{MPQRS}\tilde{F}_{N}^{\;\;\, PQRS} \\
&+& {g_s e^{-\Phi} \over 4} ( H_{MPQ}H_{N}^{\;\;\, PQ} + e^{2\Phi} \tilde{F}_{MPQ}\tilde{F}_{N}^{\;\;\, PQ})
 \\ &-& {g_s e^{- \Phi} \over 48} g_{MN}(H_{PQR}H^{PQR} + e^{2\Phi} \tilde{F}_{PQR} \tilde{F}^{PQR}),
 \nonumber
\end{eqnarray}
where $M,N = 1, ..., 10$ and $\tilde{F}_3 \equiv F_3 - C_0 H_3$.
Checking the consistency of our ansatz, we see that  the axion $C_0$ can indeed be set to zero as long as  $F_3$ and $H_3$ don't have any components which are polarized in the same
directions.  We have already satisfied the $\tilde{F}_5$ Bianchi identity, and the $F_3$  and $H_3$ Bianchi identities are trivially satisfied by our ansatz as well.  The $F_{3}$ equation of motion can be seen to be be satisfied using $\tilde{F}_{5} \wedge H_{3} = 0$ and
\begin{eqnarray}
* \, e^\Phi F_{3} = \bar{M} r^3 e^{\Phi + 4a-b} \,  {\rm vol}_{R^4} \wedge dr \wedge (e_{\phi_{1}} \wedge e_{\theta_{1}} - e_{\phi_{2}} \wedge e_{\theta_{2}}) \,,
\end{eqnarray}
which is closed.

The equations that remain are the dilaton, $H_3$, and Einstein equations.  The first two take the explicit form
\begin{eqnarray}
\label{eqn:phi} \nabla^{2} \Phi  =   - {r^{2} g_s  e^{- \Phi} \over e^{-6a + 4c} }(k'^{2} - {e^{2\Phi} \bar{M}^{2} \over r^{2}e^{2b}})  \,,
\end{eqnarray}
\begin{eqnarray}
\label{eqn:k}
{\partial \over \partial r} \left( r^{5} e^{-\Phi} k' e^{4a + b}  \right)  = g_s  \bar{M} r^{3} \left({ N_{eff} \over  e^{-8a + b + 4c}} \right) \,,
 \end{eqnarray}
 In addition, we have three distinct Einstein equations, which are
\begin{eqnarray}
\label{eqn:xxeinstein}
R_{xx} =   - {g_s^2 r^{2} \over 4} {\bar{N}_{eff}^2 \over e^{-12a + 2b + 8c}} - {g_s e^{-\Phi} \over 4 e^{-8a + 4c}}( r^{4} k'^{2} + {r^{2 }e^{2 \Phi} \bar{M}^{2} \over  e^{2b}}) + {2p \kappa_{10}^{2} r^{2}T_{3} \delta(r-r_{0}) \over 2 e^{-2a}\sqrt{g_{6}}},
\end{eqnarray}
for the $xx$ component,
\begin{eqnarray}
\label{eqn:ppeinstein}
R_{\psi \psi} =  {g_s^2 \over 36} {\bar{N}_{eff}^2 \over  e^{-8a + 8c}} - {g_s e^{- \Phi} e^{2b} \over 36 e^{-4a + 4c}}( r^{2} k'^{2} - {3 e^{2\Phi} \bar{M}^{2} \over e^{2b} }) -  {2p \kappa_{10}^{2} T_{3} e^{2b} \delta(r-r_{0})\over 18 e^{2a}\sqrt{g_{6}}},
\end{eqnarray}
for the $\psi \psi$ component, and
\begin{eqnarray}
\label{eqn:tteinstein}
R_{\theta \theta} =  {g_s^2 \over 24} {\bar{N}_{eff}^2 \over e^{-8a + 2b + 6c}} + {g_s e^{-\Phi} \over 24 e^{-4a + 2c}}( r^{2} k'^{2} + {e^{2 \Phi} \bar{M}^{2} \over  e^{2b}}) - {2p \kappa_{10}^{2} T_{3} e^{2c} \delta(r-r_{0})\over 12 e^{2a} \sqrt{g_{6}}},
\end{eqnarray}
for either $\theta_{i}\theta_{i}$ component;  the full expression for the components of the Ricci tensor for the metric (\ref{eqn:metric}) can be found in Appendix C. 

We can gain some intuition concerning the way the various equations couple with one another.
One may see that the RHS of the $xx$-Einstein equation  (\ref{eqn:xxeinstein}) acts as a source for $e^{4a}$.  Substituting this result into the other two Einstein equations (\ref{eqn:ppeinstein}), (\ref{eqn:tteinstein}), one can show that $e^{2c}$ is not sourced and it is consistent to set $c=0$.

However, in general the function $b$ is nontrivial.  This is interesting because $e^{b-c}$ represents the relative warping between two factors of $T^{1,1}$; the four dimensional space comprised by the $S^{2}$s on the one hand, and the $\psi$-circle acted on by $U(1)_\psi$ on the other.
Thus if $c$ is trivial, but $b$ is not, then the compact space is {\em squashed}.

In fact, we find that the squashing $e^{2b}$ is sourced by the combination of fluxes,
\begin{eqnarray}
\label{eqn:squashing}
- { e^{- \Phi} \over  e^{-4a}}( k'^{2} - {e^{2\Phi}\bar{M}^{2} \over r^{2}}) .
\end{eqnarray}
Observe that the source for the squashing term (\ref{eqn:squashing}) contains the same function as the right-hand-side of the dilaton equation (\ref{eqn:phi}), which referring to (\ref{OrigDilatonEqn}) can be shown to be proportional to the flux combination 
\begin{eqnarray}
\tilde{F}_3 \cdot \tilde{F}_3 - e^{-2 \Phi} H_3 \cdot H_3\quad \propto \quad {\rm Re}\ G_+ \cdot G_- \,,
\end{eqnarray}
where as usual we have defined the complex flux $G_3$ and its imaginary-self-dual and imaginary-anti-self-dual combinations:
\begin{eqnarray}
G_3 \equiv F_3 - \tau H_3 \,, \quad \quad G_\pm \equiv i G_3 \pm \ast_6 G_3 \,.
\end{eqnarray}
The KT background has flux that is purely imaginary-self-dual, $G_- = 0$.  Thus we find that a perturbation that generates an imaginary anti-self-dual flux in addition must both source the dilaton, and squash the compact space.

We shall find that exactly this occurs.  The antibrane itself does not directly source the flux, but sources the metric warp factor $e^{2a}$.
This in turn acts as a source in (\ref{eqn:k}) and produces a nontrivial perturbation to $k$ corresponding to flux that is no longer imaginary-self-dual.  Finally, this flux causes the dilaton to run and the $T^{1,1}$ to squash.  We thus find the interesting result that even branes smeared so as to break no symmetries of $T^{1,1}$ in the UV still end up modifying its shape.  The compact space is highly symmetric, but not symmetric enough to prevent this.

\subsection{Linearized solution and its properties}

We now explicitly solve equations (\ref{eqn:phi})--(\ref{eqn:tteinstein}) to linear order.  To do so, we expand the functions, $e^{2a}, e^{2b}, e^{2c}, \Phi, k$,  in a power series around infinity at $1/r = 0$.
We consider an ansatz where each function has corrections at order $1/r^4$ as well as $\log r /r^4$; higher order terms are neglected and the coefficient of the perturbation is kept only to linear order.  We find a consistent solution to the linearized equations depending on two parameters, which we call ${\cal S}$ and $\phi$:
\begin{eqnarray}
\label{aFluctSoln}
e^{-4a} &=& {1 \over 4} g_s \bar{N}+ {1 \over 8} (g_s \bar{M})^2  +  {1 \over 2} (g_{s}\bar{M})^2 \log r \\
\nonumber &&+ {1 \over r^4} \left[  \left( {1 \over 32} g_s \bar{N} + {13 \over 64} (g_s \bar{M})^2+  {1\over 4} (g_{s}\bar{M})^2  \log r \right) {\cal S} - {1 \over 16} (g_{s} \bar{M})^2\, \phi   \right]  \,, \\
e^{2b} &=& 1 + {1 \over r^4}\,  {\cal S} \,, \\
e^{2c} & = & 1 \,, \\
\label{kFluctSoln}
k &=& g_{s}\bar{M}  \log r + {1 \over r^4} \left[ \left({ 3 \over 8}  {\bar{N} \over \bar{M}} + {11\over 16}  g_{s}\bar{M}  + { 3\over 2} \, g_{s}\bar{M}   \log r \right) {\cal S} - {1 \over 4}  g_{s}\bar{M} \,\phi \right] \,, \\
\Phi &=& \log g_{s} + {1 \over r^4}  \left[ \phi  - 3 \,{\cal S} \log r \right] \,,
\label{PhiFluctSoln}
\end{eqnarray}
We claim that this is in fact the unique solution consistent with our ansatz.  We provide additional justification for this by identifying this solution as a particular member of a class of solutions generated by Aharony, Buchel and Yarom \cite{ABY}, which are claimed to exhaust the solutions to a class of perturbations around KT broad enough to include our ansatz; the comparison to the notation of \cite{ABY} is made in Appendix D.

We would like to calculate the total mass of this solution.  Since the cascade geometry is asymptotic neither to flat space nor to AdS, we cannot use ordinary expressions for the mass specialized to those geometries.  In Appendix A we review the Hawking-Horowitz mass \cite{HH} which can be computed
in more general spaces.  We find for the contribution of the perturbed background to the normalized extrinsic curvature integral,
\begin{eqnarray}
\nonumber
{\cal N} \int {\cal K} &=& 3 r^4 + {(g_s \bar{M})^2 r^4 \over 4}e^{4a_0}  - {1 \over 2}{\cal S}\\
&-&  e^{4a_0} \left[ \left( {1 \over 16} g_s \bar{N}  + {5 \over 32} (g_s \bar{M})^2 + {1 \over 2} (g_s \bar{M})^2 \log r \right) {\cal S}- {1 \over 8} (g_s \bar{M})^2 \phi \right] \\
&-& e^{8a_0} {(g_s \bar{M})^2 \over 4} \left[ \left( {1 \over 32} g_s \bar{N} + {13 \over 64} (g_s \bar{M})^2 + {1 \over 4} (g_s \bar{M})^2 \log r \right) {\cal S} - {1 \over 16} (g_s \bar{M})^2 \phi  \right] \,,
\nonumber
\end{eqnarray}
where we have suppressed a volume factor $V_3\, {\rm vol}(T^{1,1})$, and $a_0$ is the background value of the warp factor,
\begin{eqnarray}
e^{-4a_0}= {1 \over 4} g_s \bar{N} + {1 \over 8} (g_s \bar{M})^2 + {1 \over 2} (g_s \bar{M})^2 \log r \,.
\end{eqnarray}
Subtracting the background value ${\cal N} \int {\cal K}_0$ for the unperturbed cascade removes the first two terms which are divergent at large $r$.   We then take the $r \to \infty$ limit; the warp factor $e^{-4a_0}$ diverges as $\log r$.
Most terms vanish, and we find the finite mass (restoring volume factors)
\begin{eqnarray}
\label{eq:mass}
E = - {1 \over 8 \pi G_{10}}\,  {\cal N} \int ({\cal K} - {\cal K}_0)  =   {3 \over 16 \pi G_{10}}\, V_3\, {\rm vol}(T^{1,1}) \, {\cal S} \,,
\end{eqnarray}
where $G_{10} = \kappa_{10}^2/ 8\pi$ is Newton's constant in ten dimensions.  We see that the mass is directly proportional to the ${\cal S}$ perturbation, while the $\phi$ perturbation is not involved at all.  Meanwhile, the value of the D3 charge is trivial due to the definition of $\tilde{F}_5$; it is simply $N_{eff}(r)$.  All corrections to $k(r)$ vanish as a power $1/r^4$ in the $r \to \infty$ limit, so this solution does not change the asymptotic charge of the cascade.

We identify the coefficient ${\cal S}$ as the correct parameter associated to the brane/antibrane perturbation.  The mode $\phi$ can be seen to reduce in the AdS limit ($M \to 0$) to the usual independent fluctuation of the dilaton, related to the ability to turn on a vacuum expectation value for the dual operator ${\cal O}_+$ in the absence of 3-form fluxes.  As the brane/antibrane probes do not source the dilaton directly, we set this parameter to vanish, as we implicitly did in the AdS case: $\phi  =0$.  Note, however, that although this means there is no non-logarithmic perturbation to the dilaton field, we are still left with a dilaton perturbation proportional to $(\log r/r^4) \, {\cal S}$.  This is because, as was emphasized previously, the perturbation sources IASD flux, which in turn squashes the $T^{1,1}$ and forces the dilaton to run.  This logarithmic perturbation is that running.

The numerical value of ${\cal S}$ is determined by the boundary conditions on the fields at the IR tip of the geometry.  Although we do not solve this precisely because we choose to avoid the details of the IR part of the geometry, we may still use the way a source contributes to the field equations to determine the dependence of ${\cal S}$ on the various parameters.
The Einstein equation (\ref{eqn:xxeinstein}) in the presence of the $p$ D3/$\D3$ sources localized at $r=r_0$ gives
\begin{eqnarray}
\partial_{r} \partial_{r} e^{-4a} + \ldots= 4 p \kappa^{2}_{10} T_{3} r_{0}^{4} {\delta(r-r_{0}) \over \sqrt{\tilde{g}_{6}}},
\end{eqnarray}  
where $\tilde{g}_{6} = dr^{2} + r^{2} ds^{2}_{T^{1,1}}$ is the unwarped metric on the transverse six-space and we have only written the second derivative term on the warp factor.  Multiplying both sides by $1/r^{5}$ and integrating over the transverse six space, we find ${\cal S} \sim (p/N) \, r_{0}^{4}$.\footnote{Strictly speaking the quantity in the denominator of $p$ is some combination of the two background contributions to the running number of colors, $N$ and $g_s M^2$.}  Thus, ${\cal S}$ is warped by $r_{0}^{4}$ and proportional to the number $p$ of brane-antibrane pairs.

This scaling of the perturbation ${\cal S}$ enables us to understand how it behaves in the AdS limit of $M \to 0$.  Examining the form of the solution for $k(r)$ (\ref{kFluctSoln}), it naively seems  to diverge in this limit due to the term proportional to $N/M$.  However, this divergence assumes that ${\cal S}$ is kept fixed in that limit; in fact it is not.  The IR scale of the throat $r_0$ depends exponentially on the fluxes as $r_0 \sim \exp (- 2 \pi N/ 3 g_s M^2)$, thus receding to zero in the AdS limit, and likewise
\begin{eqnarray}
\label{SScale}
{\cal S} \sim{ p \over N} \, e^{- 8 \pi N / 3 g_s M^2} \,,
\end{eqnarray}
sending the total perturbation to zero as $M \to 0$.  This confirms explicitly that the supersymmetry breaking is exponentially small relative to the fundamental scale, corresponding instead to the IR scale of the bottom of the throat.

This behavior reflects the fact that although the KT solution does not directly include the smooth IR tip, nonetheless the KS deformation is the only way to resolve it in a non-singular fashion, and hence perturbations around KT ``know" that the IR tip is their only consistent resolution.  The geometry is aware that it cannot support a solution of the kind we have found here once the AdS limit is taken --- indeed we saw no trace of it in the analysis of section \ref{ConformalSec} --- and the fact that it must be scaled to zero in that limit corresponds to the retreat of the confining scale down to zero.

\subsection{Brane/antibrane force}

Another check on our linearized solution is provided by a comparison to the calculation in \cite{KKLMMT} of the force between a well-separated brane/antibrane pair in a warped throat. (In \cite{KKLMMT} the calculation was approximated assuming a background AdS geometry; the true conifold case was considered in \cite{GangOfSix}).  Placing the antibrane at the cutoff tip at $r_{0}$ and the brane at $r$, for $r \gg r_{0}$, the attractive potential between the two was found to be
\begin{eqnarray}
V(r) = - {2 T_{3} r_{0}^{8} \over N R^{4} r^{4}} \, .
\end{eqnarray}
Also, observe the warping proportional to $r_{0}^{8}$.  This potential results in an attractive force $F(r)$ of magnitude
\begin{eqnarray}
\label{BraneForce}
|F(r)| = {8T_{3}r_{0}^{8} \over N R^{4} r^{5}} \,.
\end{eqnarray}
To compare with this result, 
we place a probe D3-brane in the background of our linearized solution, which already includes the effects of $p$ $\D3$-branes at $r = r_0$, and investigate the resulting force.  To lowest order in $\alpha'$, the brane probe is governed by a Lagrangian consisting of the DBI and CS terms
\begin{eqnarray}
S_{D3} = - T_{3} \int d^{4}x\ r^{4} e^{4a}\sqrt{1 + {e^{-4a} \over r^{4}} \partial_{\mu} r \partial^{\mu}r} + T_{3} \int d^{4}x \, \alpha \,,
\end{eqnarray} 
where $\mu = 0, ..., 3$.

Expanding the action, we find in fact that there is {\em zero} force on the probe brane at the order
to which we have worked.  In fact this is the result of a precise cancellation; if any of the coefficients in the perturbation (\ref{aFluctSoln})-(\ref{PhiFluctSoln}) are tuned away from their actual precise numerical values (which involve numbers like $13/64$ and $11/16$), then a force would result of order $F(r) \sim r_0^5/ R^4 r$, which falls off much more slowly than (\ref{BraneForce}) and would have dominated over that result at large $r$.  Only this cancellation prevents the cascade geometry from having a much stronger flux-aided brane/antibrane force than the result of \cite{KKLMMT}.

Given then that the expected force on a probe brane is $\emph{not}$ the result of the leading dimension four operator perturbations that we found, one may ask whether this force exists and what produces it.   Recall that we have only worked to order $1/r^4$, and hence perturbations vanishing faster at large $r$ will be invisible.  If one postulates the existence of additional terms at order $1/r^n$  with $n \geq 4$ and coefficient ${\cal F}$ (including possible log terms), but keeps both ${\cal S}$ and ${\cal F}$ only to linear order, then one must have ${\cal F} \sim p \, r_0^n$ and a force is generated,
\begin{eqnarray}
\label{eqn:forcegeneral}
|F(r)| \sim T_{3}{p \, r_{0}^{n} (\delta + \epsilon \log r) \over  r^{n - 3} N_{eff}^{2}(r)},
\end{eqnarray}
for some constants $\delta, \epsilon$ fixed by the equations of motion.  The correct force law is seen to arise for $n=8$.  

There are two modes consistent with the symmetries of our ansatz corresponding to higher-dimension operators that could be turned on in the background at higher order: a dimension six operator ${\cal O}_6$ associated with the warp factors $e^b$ and $e^c$, and the familiar ${\cal O}_8$ operator discussed in section~\ref{ConformalSec}.  Given the matching of the force calculation with $n=8$, it is natural to surmise that the operator ${\cal O}_8$
 is turned on in the background as well at order $p$, and is responsible for the brane/antibrane force; it should be visible if one were to include perturbations out to $1/r^8$.  The operator ${\cal O}_6$ would then either be absent, or would participate in another nontrivial cancellation to avoid producing a $1/r^3$ contribution to the force (\ref{eqn:forcegeneral}).  (Note that the ${\cal S}$ perturbation would also correct the $1/r^5$ force law at order $p^2$.)

Due to the sensitivity of the $1/r$ force cancellation to the precise coefficients of the solution, the above calculation is a highly nontrivial check on our result; although the dimension eight operator is responsible for the force on a probe brane, we see this is only possible with the precise cancellation that occurs with terms associated to the dimension four operator.

\subsection{Three-form flux type}
\label{sec:flux}

As mentioned before, the KT (and KS) geometries possess imaginary-self dual flux $G_+$, with $G_- = 0$.  Moreover, this flux is known to be of type $(2,1)$ relative to the complex structure of the geometry; in general ISD flux can be $(2, 1)$ or $(0,3)$ while IASD flux is $(3,0)$ or $(1,2)$.  In this subsection we determine the Hodge or index type of the flux perturbation.

The complex flux $G_3$ takes the form
\begin{eqnarray}
\label{G3Eqn}
G_3 \equiv F_3 - i e^{-\Phi} H_3 = \left( \bar{M} e_\psi - i e^{-\Phi} \partial_r k \, dr \right) \wedge {\omega_2 \over 3} \,,
\end{eqnarray}
where $\omega_2 \equiv 3(e_{\theta_{1}}\wedge e_{\phi_{1}} - e_{\theta_{2}}\wedge e_{\phi_{2}})$.
Tensors associated to different Hodge types in this background were classified by Herzog, Klebanov and Ouyang \cite{HKO}.  They provided a dictionary to translate the forms $dr$, $e_\psi$ and $\omega_2$ we use into the four complex coordinates $z_i$ defining the deformed conifold via $\sum_{i=1}^4 z_i^2 = \epsilon^2$:
\begin{eqnarray}
 \frac{d r}{r} + i e_{\psi} &=& {2 \over 3} \frac{1}{r^{3}} \bar{z}_i d z_i \,, \quad \quad
 \frac{d r}{r} - i e_{\psi} = {2 \over 3} \frac{1}{r^{3}} z_i d \bar{z}_i \,, \\
\omega_{2}  &=&  {i \over r^{6}} \epsilon_{ijkl} z_{i}\bar{z}_{j} dz_{k} \wedge d\bar{z}_{l} \,,
\end{eqnarray}
in terms of which the holomorphic structure is manifest; note the scaling $|z|^2 \sim r^3$.
Then for (\ref{G3Eqn}) we have,
\begin{eqnarray}
G_3 =  {1 \over 9 r^9} \left[ \left(\bar{M} + e^{-\Phi} r \partial_r k \right)\bar{z}_m d z_m    -  \left(\bar{M} - e^{-\Phi} r \partial_r k \right) z_m d \bar{z}_m \right]  \epsilon_{ijkl}  z_{i}\bar{z}_{j} dz_{k} \wedge d\bar{z}_{l} \,.
\end{eqnarray}
We readily see that the only flux types are a $(2,1)$ part (hence ISD) proportional to $(\bar{M} + e^{-\Phi} r \partial_r k)$, and a $(1,2)$ part (hence IASD) proportional to $(\bar{M} - e^{-\Phi} r \partial_r k)$.  The product of these quantities is the product of ISD and IASD fluxes, and indeed is proportional to the flux combination (\ref{eqn:squashing}) sourcing the dilaton and the squashing, which we previously identified as Re $G_+ G_-$.

Since the $\bar{M}$ piece of each $G_+$ and $G_-$ is part of the background, the contribution of the fluctuation comes entirely from the $e^{-\Phi} r \partial_r k$ terms, and hence the brane backreaction contributes fluctuations of $(2, 1)$ and $(1, 2)$ fluxes with equal magnitude.  For completeness, the fluxes are
\begin{eqnarray}
G_{+,0} & = & {2 i \over 9 r^{9}}\left[ 2 \bar{M} - {{\cal S}  \over r^{4}}\left({3 \over 2}{\bar{N} \over g_{s} \bar{M}} + {5 \over 4}\bar{M} + 3 \bar{M} \log r \right) \right](\epsilon_{ijkl} z_{i} \bar{z}_{j} \bar{z}_{m} dz_{k} \wedge d\bar{z}_{l} \wedge dz_{m}) \,, \\
G_{-,0} & = & {2 i \over 9 r^9} { {\cal S} \over r^4 }\left({3 \over 2}{\bar{N} \over g_{s} \bar{M}} + {5 \over 4}\bar{M} + 3 \bar{M} \log r \right) (\epsilon_{ijkl} z_{i} \bar{z}_{j} z_{m} dz_{k} \wedge d\bar{z}_{l} \wedge d\bar{z}_{m}) \,.
\end{eqnarray}
We emphasize that the above represents the decomposition of the perturbed flux into the Hodge types of the {\em background} geometry, denoted by the $0$ subscript.  However, the squashing factor $e^b$ will also modify the metric, and so a natural question is whether the perturbed flux is ISD with respect to the {\em perturbed} geometry.  One may show that the fluxes take the form,
\begin{eqnarray}
G_\pm  =  \left[\bar{M} \pm e^b e^{-\Phi} r \partial_r k \right]  \left( i e_\psi \pm e^{-b} {dr \over r} \right) \wedge {\omega_2 \over 3} \,,
\end{eqnarray}
in the modified background, which evaluates to
\begin{eqnarray}
G_{+} & = & \left[ 2 \bar{M} - {{\cal S}  \over r^{4}}\left({3 \over 2}{\bar{N} \over g_{s} \bar{M}} + {3 \over 4}\bar{M} + 3 \bar{M} \log r \right) \right] \left( i e_\psi + e^{-b} {dr\over r} \right) \wedge {\omega_2 \over 3} \,, \\
G_{-} & = & { {\cal S} \over r^4 }\left({3 \over 2}{\bar{N} \over g_{s} \bar{M}} + {3 \over 4}\bar{M} + 3 \bar{M} \log r \right)  \left( i e_\psi - e^{-b} {dr\over r} \right) \wedge {\omega_2 \over 3} \,,
\end{eqnarray}
and hence we see the perturbed flux is not ISD (or IASD) with respect to the perturbed geometry, either.

We note in passing that the leading corrections to the KT solution coming from the full KS solution, indicative of chiral symmetry breaking, actually fall off less quickly than our perturbation far from the tip, as $G_3 \sim M (r_0/r)^3$ \cite{Loewy}.  Hence our perturbation is not the leading correction to the KT geometry; it is, however, the leading term proportional to $p$ and the leading term that breaks supersymmetry.  An antibrane solution valid over the entire KS geometry should contain corrections to the chiral symmetry breaking structure proportional to $p$ as well.

\section{SUSY Breaking in the Field Theory Dual}
\label{DualSec}

Because the metastable $\D3$-brane states have not proved amenable to a direct weak 't Hooft coupling field theory analysis (for work in this direction in
slightly more general geometries, see \cite{ABFK}), the characteristics of the supersymmetry breaking that we can learn from the operator vacuum expectation values using the AdS/CFT dictionary may be a useful source of insight.

Even before looking at the VEVs, we can immediately see one interesting feature.  As explained in \cite{Klebanov-Ouyang-Witten}, the anomaly-related R-symmetry breaking
in the Klebanov-Strassler solution from $U(1)$ to $Z_{2M}$ is visible through the behavior of the $C_2$ Ramond-Ramond gauge field.  
In our solution, $F_3$ and hence $C_2$ are completely unchanged from their background value.  So 
the SUSY breaking does ${\it not}$ (at least at leading order) give rise to R-breaking independent from that already present in the KT and KS solutions.  

Because the SUSY-breaking does not provide another independent order parameter for R-breaking,
we conclude that it is either D-term breaking, or F-term breaking which preserves R.  The reader should note that while
${\cal O}(1)$ R-breaking at the SUSY-breaking scale would imply F-term breaking (or at least a significant F-component
to the breaking), the absence of R-breaking is consistent with either D or F-term breaking --- with the ISS models \cite{ISS}
providing a
canonical example where F-term breaking preserves R.

To learn more about the supersymmetry breaking, let us try to characterize the state in the field theory more precisely.
Being a set of normalizable fluctuations of the supergravity fields, our solution implies a number of vacuum expectation values for the dual operators.  The techniques of holographic renormalization \cite{HoloRenorm} for calculating expectation values in an asymptotically anti-de Sitter background have been generalized to the cascade geometry by Aharony, Buchel and Yarom \cite{ABY} and our solution can be treated by their techniques; we match our notation to theirs in Appendix D.  Using their formulae, we find the expectation values (including the $\phi$ mode for completeness):
\begin{eqnarray}
\label{vevs}
\langle T_{\mu\nu} \rangle &=& - {3 \eta_{\mu \nu} \over 2} \, {\cal S} \,, \\
\langle {\cal O}_{+} \rangle &=& - 3 {\cal S} - 4 \phi \,, 
\label{OplusVEV} \\
\langle  {\cal O}_{-} \rangle &=& {12 \over \bar{M}} {\cal S} \,,
\label{OminusVEV}
\end{eqnarray}
where ${\cal O}_{+}$ represents the operator dual to $e^{-\Phi}$ while ${\cal O}_{-}$ represents the operator associated  to $e^{-\Phi} B_2$.\footnote{The imaginary parts of ${\cal O}_\pm$ operators, dual to modes of $C_0$ and $\int C_2$, are not given an expectation value in our solution.}  As with the perturbed solution itself, the vev $\langle {\cal O}_- \rangle$ seems singular in the $M \to 0$ limit at first glance, but actually goes smoothly to zero since ${\cal S}$ vanishes exponentially (\ref{SScale}).

If one defines the 5D boundary stress tensor as $T_{\mu\nu}^{5D} = T_{\mu\nu}/  8 \pi G_5$ with $G_{5} = G_{10}/{\rm vol}(T^{1,1})$ as in \cite{ABY}, one sees immediately that $ \langle T_{00}^{5D} \rangle$ exactly matches the calculation of the mass density of the spacetime presented in (\ref{eq:mass}).  The nonzero value of the energy density is confirmation that the state we are studying indeed breaks supersymmetry, and the scaling (\ref{SScale}) of ${\cal S}$ confirms that it is broken at an exponentially small scale.

The running of the 
coupling associated to ${\cal O}_-$  (\ref{runaway}) generates a quantum violation of conformal invariance proportional to the beta function:
\begin{eqnarray}
\label{ConformalAnomaly}
T^\mu_\mu = - {1 \over 2} \sum_i \beta_i {\cal O}_i =  - {1 \over 2} \beta_- {\cal O}_- \,,
\end{eqnarray}
where contributions proportional to the conformal anomaly vanish for a Minkowski background.  This operator relation must hold in any state, in particular our supersymmetry-breaking state.  It was verified by \cite{ABY} that for the beta function at hand, which becomes $\beta_- = \bar{M}$ for us,\footnote{Note we have defined ${\cal O}_-$ as directly dual to $e^{-\Phi} B_2$ with no numerical factors, which leads to the given normalization.} the corresponding relation
\begin{eqnarray}
\label{VevRelation}
 \langle T^\mu_\mu \rangle = -{1 \over 2}\bar{M}  \langle {\cal O}_- \rangle \,,
\end{eqnarray}
is indeed satisfied, as can be confirmed from (\ref{vevs}) and (\ref{OminusVEV}).  Note that the dependence of  $\langle {\cal O}_+ \rangle$ on $\phi$, the parameter we set to zero for the $\D3$-brane solution, but which indicates another set of possible states in the theory, rules out the appearance of ${\cal O}_+$ in such a relation.

Now, let us return to the question of whether the supersymmetry breaking can be characterized as D-term or F-term breaking.  For our Poincar\'e-invariant state, the trace of the stress tensor is proportional to the energy density; thus (\ref{VevRelation}) is a relation between the energy of the state and the expectation value of ${\cal O}_-$.  Hence if we can determine the Lorentz-invariant terms in ${\cal O}_-$ that can acquire a VEV, we can characterize the nature of the supersymmetry breaking.  Additionally, a supersymmetric partner of the operator relation (\ref{ConformalAnomaly}) relates the fermionic partner of ${\cal O}_-$, call it $\Psi_-$ to the supercurrent, which when supersymmetry is spontaneously broken is linear in the goldstino; thus we expect $\Psi_-$ to be linear in the goldstino when its auxiliary fields are given their VEV.

For these reasons it is interesting to understand ${\cal O}_-$ better.
We have mentioned previously that ${\cal O}_+$ and ${\cal O}_-$ can be identified with the sum and difference of the coupling constants of the two factors in the gauge group.  Properly speaking however, the story is a little more complicated \cite{KW, KS}.  In the conformal limit, there are {\em three} $SU(2)_1 \times SU(2)_2$-preserving dimension four scalar operators: the F-terms of
Tr$(W_{\alpha,1}^2)$, Tr$(W_{\alpha,2}^2)$ and Tr $(A_i B_k A_j B_l \epsilon^{ij}\epsilon^{kl})$, the first two associated to the two gauge couplings $g_1$ and $g_2$ and the last one being the superpotential term (\ref{Superpotential}), which has associated coupling $\lambda$.  There are, however, only two combinations of the three couplings that are invariant under all the anomalous global symmetries in the CFT, which we can take as  $J = (\lambda^2 \Lambda_1\Lambda_2)^N$ and $I = (\Lambda_2 / \Lambda_1)^N$ where $\Lambda_1$ and $\Lambda_2$ are the usual complexified dynamically-generated scales for the two gauge factors; it is easy to see that $J$ and $I$ do involve the sum and differences of the gauge couplings $1/g_1^2$ and $1/g_2^2$, respectively.

The freedom to choose these two parameters in the CFT corresponds to the two moduli of the theory; put differently, there is a two-complex-dimensional space of conformal fixed points in the three-dimensional space of $\Lambda_1$, $\Lambda_2$ and $\lambda$.  These can be identified with the moduli $\tau \equiv C_0 + i e^{-\Phi}$ and $\varphi \equiv  e^{-\Phi} \int_{S^2} (B_2 + i C_2)$ of the gravity dual by matching the action of the $Z_2$ center of the $SL(2,Z)$ S-duality of type IIB string theory, which on the field theory side acts as charge conjugation combined with an exchange of the two gauge factors and thus leaves $A$ and $B$ invariant \cite{KW}. This symmetry preserves both $\tau$ and $J$, which we thus match together (calling the associated operator ${\cal O}_+$), while flipping the sign of $\varphi$ and $I$, which we thus match (calling the associated operator ${\cal O}_-$).  More strictly speaking, $\tau$ and $\varphi$ are associated to $\log J$ and $\log I$, respectively.
A more detailed discussion of the identification of the $Z_2$ odd modulus with $I$, using the fact that one can obtain the KW theory from a
mass-perturbed ${\cal N}=2$ superconformal field theory, appears in \cite{Strassler}.

Away from the conformal limit, the generalizations of $J$ and $I$ are \cite{KS}:
\begin{eqnarray}
\label{IandJKS}
J \equiv \lambda^{2N +M} \Lambda_1^{3M+N} \Lambda_2^{N - 2M} \,, \quad I \equiv \lambda^{3M} { \Lambda_2^{N - 2M} \over \Lambda_1^{3M +N} } \left[ {\rm Tr} \ (A_i B_j A_k B_l) \epsilon^{ik} \epsilon^{jl} \right]^{2M} \ .
\end{eqnarray}
For nonzero $M$, we see $I$ is no longer invariant built solely out of coupling constants, but instead becomes field-dependent.  This reflects 
the fact that it is no longer associated to a modulus; the difference of the gauge couplings starts to run as in  (\ref{runaway}).

From all this we learn that in the conformal limit, ${\cal O}_-$ (unlike ${\cal O}_+$) does not involve the superpotential, and has the simple form
\begin{eqnarray}
{\cal O}_- \sim {\rm Tr} \, \left[ (F_1^2 + \lambda_1 \partial \lambda_1 + D_1^2) - (F_2^2 + \lambda_2 \partial \lambda_2 + D_2^2) \right] \,,
\end{eqnarray}
where we have schematically indicated the gauge fields, gauginos, and auxiliary $D$ fields for the two gauge factors.  Hence a Lorentz-invariant expectation value for this operator must involve solely the auxiliary fields in the gauge multiplet, and thus constitutes D-term supersymmetry breaking.

However, our metastable states only arise in the non-conformal theories with $M \neq 0$ --- there are no Lorentz-invariant metastable states
in the conformal theory.  Away from the conformal limit, we see from (\ref{IandJKS}) that the superpotential does begin to mix into ${\cal O}_-$.  Thus in principle, the auxiliary fields $F_A$ and $F_B$ may begin to contribute to an F-term supersymmetry breaking component as well.  
While the fact that in the conformal limit ${\cal O}_-$ is dominated by D-terms is suggestive of D-term supersymmetry breaking, the magnitude of
the energy density is small enough that the subdominant F-term mixing could parametrically account for the full result. 
Therefore, while
our results suggest that the breaking may be dominated by D-terms, we cannot give a precise argument that this must be so.
In fact, we note that in a strongly coupled gauge theory where one cannot identify the Goldstino uniquely as a component of some (weakly coupled) gauge or chiral multiplet, there may well be no invariant distinction between D- and F-term supersymmetry breaking.\footnote{In a theory with a U(1) factor
in the gauge group and a Fayet-Iliopoulos D-term, this situation may be different.  In coupling to supergravity,
the FI term causes shifts to gauge charges and may thus distinguish the two possibilities.  However, our theory is
an $SU(N) \times SU(N+M)$ gauge theory with no U(1) factors in the gauge group.  We thank N. Seiberg for
discussions of this possibility.}

\section{Conclusion}

Motivated by the desire to find gravity duals of vacua with exponentially small supersymmetry breaking scale, 
we studied linearized perturbations of supergravity backgrounds.  For backgrounds dual to superconformal field theories, we found a universal dimension eight operator, and a possible
identification between backgrounds perturbed by a VEV for this operator and brane/antibrane configurations, but nonetheless supersymmetry remained unbroken.

Our primary interest was to study perturbations of backgrounds dual to confining theories, concentrating on the KS solution.  We identified linearized perturbations corresponding to the metastable SUSY breaking states of \cite{KPV}.  Let us summarize the effects of the SUSY breaking $\D3$-branes on the background geometry.  The $\D3$-branes directly source the warp factor; in fact, they require at least two warp factors (or, an overall warp factor and a ``squashing'' factor).  The warp factor then acts as a source for the background three-form flux, and we find that both $G_{\pm}$ are activated.   The combination of both $G_+$ and $G_-$ flux then sources both the dilaton and the squashing factor. As a result, the
final solution differs in several precise ways from the supersymmetric solutions: the $T^{1,1}$
is squashed, the three-form flux includes (1,2) components, and the dilaton is forced to run.

There are a number of promising directions for future work.  
Soft-breaking terms induced by SUSY-breaking three-form fluxes on D3-brane probe fields
were found in \cite{Grana}.  These works assumed a single warp factor, i.e. a conformally
Calabi-Yau metric ansatz (and as a consequence, worked out soft terms in flux backgrounds that
in general do not solve the classical closed string equations of motion).  Therefore, their results cannot be literally applied to our solutions.  It would be interesting to modify these analyses of the soft terms to make them applicable to more general backgrounds like the ones we found.  
It has been suggested that antibrane SUSY breaking in warped throats is naturally sequestered,
with detailed arguments appearing in \cite{KMS}.  If this is so, one would expect the leading
approximation to the 
soft terms to vanish in many circumstances.   Additionally, the background we describe here and related configurations may have applications in phenomenologically-motivated settings like that studied in \cite{GGG}.

Compactification of supergravity solutions of the form presented here is still a nontrivial open problem.  The simplest IIB backgrounds in a compact setting have a conformally Calabi-Yau metric and no (1,2) fluxes \cite{GKP}.  It may be possible to find more general compact, classical backgrounds which allow more general supersymmetry-breaking possibilities such as the one presented here.

There are also several clear steps one could take to make our solutions more complete:

\smallskip
\noindent
$\bullet$  We worked to leading nontrivial order in $1/r$.  One could work to higher orders in $1/r$, to e.g. capture the leading force on a brane
probe at large $r$, which was approximated
in \cite{KKLMMT} (recall that we found the expected ``miraculous'' force cancellation at the order we worked).  Similarly one could
also keep terms of higher order in the $p/N$ expansion.

\smallskip
\noindent
$\bullet$ In order to simplify the analysis, we ``smeared'' the $\D3$-branes on the compact space.  This preserved the $SU(2) \times SU(2)$ isometry group of $T^{1,1}$.  There are various assumptions one could make to relax this
condition.  For instance, the probe discussion in \cite{KPV} indicates that the antibranes embiggen to wrap an
an $S^2 \subset S^{3}$ in the $T^{1,1}$.  In this circumstance, the global symmetry is reduced to $SO(3)$.
Finding solutions with this reduced symmetry may be tractable.

\smallskip
\noindent
$\bullet$ Most ambitiously, one would like to find solutions for $\D3$-branes that are valid over the whole
KS throat.  As a first step, one could find another class of perturbative solutions valid in the
``near tip" region (where the leading order metric and background fields are very simple), and then match them at some intermediate $r$ to the large $r$ solutions
we have presented in this paper.  It is only by actually finding the solutions that extend to the
IR and are built on ``real" brane sources, that one will be able to determine the correct quantization
of the parameter ${\cal S}$ that appears in the linearized solutions.  Similarly, only
by extending our solutions to the tip is it likely that we can see the perturbative instability 
found in \cite{KPV}, that
sets in if one leaves the regime $p \ll M$.

\bigskip\bigskip\bigskip

\centerline{\bf{Acknowledgements}}

\bigskip
We are grateful to Wu-yen Chuang for discussions and collaboration in the early stages of this work.  We would like to give special thanks to Igor Klebanov and Liam McAllister for discussions and helpful comments on an early draft of this paper.
We would also like to thank Ofer Aharony, Shanta de Alwis, Daniel Baumann, Michael Dine, Shesansu Pal, Michael Peskin, Nathan Seiberg, Eva Silverstein, and Marika Taylor for helpful discussions.  We thank Mustafa Amin and John Conley for providing us with their Mathematica program.  S.K. and M.M. are grateful for the kind hospitality of the University
of Colorado Theoretical High Energy Physics group during TASI 07, while
this work was in progress.
The work of O.D. was supported by the DOE under grant DE-FG02-91-ER-40672.
The work of S.K. and M.M. was supported in part by NSF grant PHY-0244728 and in part by the
DOE under contract DE-AC03-76SF00515.

%\newpage

\section*{Appendix A: An expression for total mass}

For spacetimes asymptoting to flat space, one may define the ADM mass using the falloff of the metric at infinity; a similar quantity exists for perturbations of  AdS space.  For the Klebanov-Strassler cascade solution which asymptotes to neither, one needs a more general definition of the mass of a spacetime. We make use of a quantity formulated in more general spacetimes by Hawking and Horowitz \cite{HH}, which reduces to the familiar cases for flat space and AdS space.  Another useful reference is \cite{HorowitzMyers}.

For the total mass of a spacetime with a timelike Killing vector, we have 
\begin{eqnarray}
E \equiv - {1 \over 8 \pi G_{D}} \int {\cal N} ({\cal K} - {\cal K}_0) \,,
\end{eqnarray}
where the norm of the timelike Killing vector is $-{\cal N}^2$, ${\cal K}$ is the trace of the extrinsic curvature and ${\cal K}_0$ is the extrinsic curvature of the zero-mass spacetime with the same asymptotic behavior and $G_{D}$ is the D-dimensional Newton constant.  The integral is taken at a fixed time, over a surface at spatial infinity; the extrinsic curvature represents the embedding of this surface in the fixed-time space. For a $D$-dimensional spacetime metric  of the form
\begin{eqnarray}
\label{MetricForm}
ds^2 = - {\cal N}(r) dt^2 + f(r) dr^2 + ds^2_{D-2}(r, \Omega) \,,
\end{eqnarray}
where the $D-2$-dimensional metric can depend on other variables $\Omega$, we may calculate the integral of ${\cal K}$ as follows.  The $D-1$-dimensional space at constant $t$ is foliated by constant-$r$ slices; the normalized vector orthogonal to these surfaces is called $n^M$ and is
\begin{eqnarray}
n^r = f^{-1/2} \,, \quad \quad n^M = 0 \,, \; {M \neq r} \,,
\end{eqnarray}
and one may check that this vector field generates a set of geodesics.  In these circumstances, the (trace of the) extrinsic curvature can be written ${\cal K} = \nabla_M n^M$, and we then can show 
\begin{eqnarray}
\int d^{D-2}\Omega \sqrt{g_{D-1}} \, {\cal K}
=  \int d^{D-2}\Omega\,  n^M \partial_M (\sqrt{g_{D-2}}) = n^M \partial_M A \,,
\end{eqnarray}
with $A$ the $D-2$-dimensional area,
\begin{eqnarray}
A \equiv  \int d^{D-2}\Omega \, \sqrt{g_{D-2}} \,.
\end{eqnarray}
To calculate $E$, the radius $r$ of the slice must be taken to infinity.

\section*{Appendix B: Flat space non-BPS brane solutions}

We would like to be able to provide more evidence that the mode (\ref{Delta8}) does indeed correspond in some way to a brane/antibrane perturbation.  We can do this by matching to a set of known solutions for a stack of branes and antibranes in flat space, and taking an appropriate near-horizon limit.  In the process we will gain a better understanding of the vanishing of the ADM mass in AdS space.

Brax, Mandal and Oz constructed a set of solutions for non-BPS, Poincar\'e-invariant branes embedded in flat space in various dimensions.  For the 3-brane case, a class of solutions is
\begin{eqnarray}
\label{BMOSolns}
e^{-4A} &=& \cosh(k\,  h(r)) - c_2 \sinh (k \, h(r)) \,, \quad \quad e^{4B} = e^{-4A} f_- f_+ \,,\\  \alpha&=& - \eta (c_2^2 - 1)^{1/2} e^{4A} \sinh (k \, h(r)) \,,
\end{eqnarray}
where $k = \sqrt{5/2}$ and
\begin{eqnarray}
f_\pm \equiv 1 \pm \left(\tilde{r} \over r \right)^4 \,, \quad \quad h(r) = \log \left(f_- \over f_+ \right) \ .
\end{eqnarray}
Here we substituted $\tilde{r}$ for BMO's $r_0$ to avoid confusion with the quantity we use in section~\ref{CascadeSec}.
These solutions correspond to a combination of D3-branes and $\D3$-branes sitting at $r=0$ but asymptotic to flat space.  We have set to zero a parameter $c_1$ in \cite{BMO} which corresponds  to a decoupled dilaton mode; since the D3- and $\D3$-branes  don't source the dilaton we chose to set its independent modes to zero.

Calculating the ADM mass and integral of $\tilde{F}_5$ and normalizing them to the BPS case of $\tilde{N}$ D3-branes,
\begin{eqnarray}
\label{BPSbranes}
e^{-4A} = e^{4B} = \alpha^{-1} =1 + {\tilde{R}^4 \over r^4} \,, \quad \quad \tilde{R}^4 = 4 \pi g_s \tilde{N} \,,
\end{eqnarray}
we find we can identify the BMO solutions (\ref{BMOSolns}) as corresponding to $\tilde{N}+\tilde{p}$ D3-branes and $\tilde{p}$ $\D3$-branes with
\begin{eqnarray}
2 \eta \tilde{r}^4 k \sqrt{c_2^2 - 1} = \tilde{R}^4 \equiv 4 \pi g_s \tilde{N} \,, \quad \quad 2 \tilde{r}^4 k c_2 = 4 \pi g_s (\tilde{N} + 2 \tilde{p}) \,.
\end{eqnarray}
Thus the parameters of the solution $\tilde{r}$ and $c_2$ can be traded for the total D3 charge and total mass; we put tildes on $\tilde{N}$, $\tilde{p}$ because they do not yet coincide with the parameters $N$, $p$ from our AdS solutions.  We see that the BPS limit of $\tilde{p} \to 0$, $\tilde{N}$ fixed is
\begin{eqnarray}
c_2 \to \infty \,, \quad \tilde{r}^4 \to 0 \,, \quad 2 k \tilde{r}^4 \sqrt{c_2^2-1} = \tilde{R}^4 \ {\rm fixed} \,,
\end{eqnarray}
and the expansion in $\tilde{p}/\tilde{N}$ is equivalent to an expansion in $1/4c_2^2$:
\begin{eqnarray}
{\tilde{p} \over \tilde{N}} = {c_2 - \sqrt{c_2^2 - 1} \over 2 \sqrt{c_2^2 - 1}} = {1 \over 4 c_2^2} + {3 \over 16 c_2^4} + \ldots \,.
\end{eqnarray}
Hence the solution can be rewritten in terms of the  scale $\tilde{R}^4 = 4 \pi g_s \tilde{N}$ associated to the background $\tilde{N}$ D3-branes and the small parameter $\tilde{p}/\tilde{N}$ associated to the $\tilde{p}$ perturbing D3/$\D3$-pairs:
\begin{eqnarray}
\label{BMOSeries}
e^{-4A} &=& 1 + (1 + {2\tilde{p} \over \tilde{N}}) {\tilde{R}^4 \over r^4} +  {2 \tilde{p} \over \tilde{N}} {\tilde{R}^8 \over r^8} + {4 \over 5} {\tilde{p} \over \tilde{N}} {\tilde{R}^{12} \over r^{12}} + {\cal O}({\tilde{p}^2 \over \tilde{N}^2}) \,, \\
e^{4B} &=& 1 + (1 + {2\tilde{p} \over \tilde{N}}) {\tilde{R}^4 \over r^4} +  {8 \over 5} {\tilde{p} \over \tilde{N}} {\tilde{R}^8 \over r^8} + {2 \over 5} {\tilde{p} \over \tilde{N}} {\tilde{R}^{12} \over r^{12}} + {\cal O}({\tilde{p}^2 \over \tilde{N}^2}) \,.
\end{eqnarray}
This is an expansion in $\tilde{p}/\tilde{N}$, though it would break down for $r/\tilde{R}$ small with $\tilde{p}/\tilde{N}$ fixed.
The $1/r^8$ term seems to be a $1/r^4$ correction to the $\tilde{R}^4/r^4$ warp factor.  This would correspond precisely to the dimension-four operator vev like $\langle T^\mu_\mu\rangle$ which we did {\em not} see in the solution of section~\ref{ConformalSec}.  However, these solutions for D3/$\D3$-pairs perturbing a stack of D3-branes do not precisely correspond to the solution we are interested in, since they asymptote to flat space instead of AdS space.

We can take an appropriate limit of the original solutions (\ref{BMOSolns}) to achieve this, however, essentially corresponding to letting $c_2$ go to infinity such that the $\cosh$ term is dropped.  Formally we can achieve this by rescaling the coordinates $x^\mu$ and $r$ such that $e^{-4A}$ and $e^{4B}$ acquire overall factors:
\begin{eqnarray}
e^{-4A} &=& Z [\cosh(k\,  h(r)) - c_2 \sinh (k \, h(r))] \,, \quad \quad e^{4B} = e^{-4A} f_- f_+ \,,
\end{eqnarray}
and then scaling the constant $Z \to 0$ with $Z c_2$ fixed.  This has the form of a perturbed AdS solution where the solution (\ref{BMOSeries}) becomes
\begin{eqnarray}
\label{BMOSeriesLimit}
e^{-4A} &=&{R^4 \over r^4} +  {4 \over 5} {p \over N} {R^{12} \over r^{12}} + {\cal O}({p^2 \over N^2}) \,, \\
e^{4B} &=&  {R^4 \over r^4} +  {2 \over 5} {p \over N} {R^{12} \over r^{12}} + {\cal O}({p^2 \over N^2}) \,.
\end{eqnarray}
Here we defined $R^4 \equiv Z \tilde{R}^4$ and $p/N \equiv (\tilde{p}/\tilde{N})/Z^2$ with $R^4$ and $p/N$ constant as $Z \to 0$.
We notice immediately that the $1/r^8$ correction disappears along with the overall ``1" in the near-horizon limit; there is no way to take the limit so as to preserve this correction.  Moreover, it is straightforward to show that the perturbations in (\ref{BMOSeriesLimit}) are of the form 
\begin{eqnarray}
\delta B = B_0 r^{-8} \,, \quad \quad \delta A = -2 \delta B \,,
\end{eqnarray}
which is precisely the perturbation we calculated in section~\ref{ConformalSec}, with
\begin{eqnarray}
B_0 = {R^8 \over 10} {p \over N} \,.
\label{FluctCoeff}
\end{eqnarray}
Thus we see that the near-horizon limit of the solution around a stack of D3-branes perturbed by a small number of D3/$\D3$ pairs is precisely the ${\cal O}_8$ mode we already identified, with the coefficient (\ref{FluctCoeff}), as long as a suitable scaling limit is taken.  This confirms the interpretation of the ${\cal O}_8$ solution as being (a limit of) a set of perturbing antibranes all the way down the $AdS$ throat.  Moreover the scaling limit to get to AdS sends the number of antibranes present in the flat space solution to zero, explaining the vanishing of the total mass of the AdS solution.

\section*{Appendix C: Ricci Tensor}

In this appendix we collect expressions for the components of the Ricci tensor $R_{MN}$ and Ricci scalar $R$ for the metric (\ref{eqn:metric}):
\begin{eqnarray}
R_{rr} & = & -{4 \over r^{2}} - 8 a'^{2} - {b' \over r} - b'^{2} - {4 c' \over r} - 4 c'^{2} + a'(- {11 \over r} + b' + 4 c') + a'' - b'' - 4 c'', \\
R_{\mu \nu} & = & - \eta_{\mu \nu} r^{2} e^{4 a}\left(4 + r b' + 4 r c' + r a'(5 + r b' + 4 r c') + r^{2}a''\right), \\
R_{\psi a} & = & e^{2a} g_{\psi a}\left(4 e^{2b - 4c} - r^{2} b'^{2} - rb'(5 + 4 r c') + r a'(5 + r b' + 4 r c') + r^{2} a'' - r^{2} b''\right), \\
R_{\phi_{1} \phi_{2}} & = & e^{2a} g_{\phi_{1} \phi_{2}}\left(4 e^{2b - 4c} - r^{2} b'^{2} - rb'(5 + 4 r c') + r a'(5 + r b' + 4 r c') + r^{2} a'' - r^{2} b''\right), \\
R_{\phi_{i} \phi_{i}} & = & {e^{-4 c} \over 18}(8 e^{4b} \cos^{2}(\theta_{i}) + 18 e^{4c} \sin^{2}(\theta_{i}) - 6 e^{2b + 2c} \sin^{2}(\theta_{i}) - 2 e^{2b + 4c} r^{2} \cos^{2}(\theta_{i}) b'^{2} \nonumber \\
& & - 15 e^{6c} r \sin^{2}(\theta_{i})c' - 12 e^{6c}r^{2}\sin^{2}(\theta_{i})c'^{2} + e^{4c}r(2 e^{2b}\cos^{2}(\theta_{i}) + 3 e^{2c} \sin^{2}(\theta_{i}))a' \nonumber \\
& & (5 + r b' + 4 r c') - e^{4c} r b' (10 e^{2b} \cos^{2}(\theta_{i}) + r (8e^{2b}\cos^{2}(\theta_{i}) + 3 e^{2c}\sin^{2}(\theta_{i})) c') \nonumber \\
& & + 2 e^{2b + 4c} r^{2} \cos^{2}(\theta_{i}) a'' + 3 e^{6c} r^{2} \sin^{2}(\theta_{i})a'' - 2 e^{2b + 4c} r^{2}\cos^{2}(\theta_{i}) b'' \nonumber \\
& & - 3 e^{6c} r^{2} \sin^{2}(\theta_{i})c''), \\
R_{\theta_{i} \theta_{i}} & = & {e^{-2c} \over 6}(-2 e^{2b} + 6 e^{2c} - e^{4c}r(5 + r b')c' - 4 e^{4c}r^{2} c'^{2} + e^{4c}r a'(5 + r b' + 4r c') \nonumber \\
& & + e^{4c}r^{2} a'' - e^{4c} r^{2} c'') \,, \\
R &=& - 2 e^{2a - 4 c}(2 e^{2b} - 12 e^{2c} + 10 e^{4c} + 4 e^{4c} r^{2} a'^{2} + e^{4c} r^{2} b'^{2} + 20 e^{4c} r c' + 10 e^{4c} r^{2} c'^{2} \nonumber \\
& &+ e^{4c} r b'(5 + 4 r c') - e^{4c}r a'(-3 + r b' + 4 r c') - e^{4c} r^{2} a'' + e^{4c} r^{2} b'' + 4 e^{4c} r^{2} c'').
\end{eqnarray}
where $a = \psi, \phi_{i}$, $i=1,2$ and primes denote derivatives with respect to $r$.

\section*{Appendix D: Matching to ABY Solutions}

Linearized solutions to the fluctuation equations around the KT background for an equivalent set of fields  were considered by Aharony, Buchel and Yarom \cite{ABY}.  They consider a considerably more general ansatz for these fields; one can show that our solution fits inside their framework.

We use the Einstein frame of \cite{PS}, $g_{\mu\nu}^{\rm PS} = g_s^{1/2} e^{-\Phi/2} g_{\mu\nu}^{\rm string}$, while the authors of \cite{ABY} use $g_{\mu\nu}^{\rm ABY} = e^{-\Phi/2} g_{\mu\nu}^{\rm string}$, so the metric we use here is related to theirs by $g_{\mu\nu} = g_s^{1/2} g_{\mu\nu}^{\rm ABY}$.  The radial variable $r$ is given in terms of their $\rho$ by $r = 1/\rho$.
The various functions of the solution are related as follows, with our notation on the left:
\begin{eqnarray}
e^{-4a} \;\; \leftrightarrow \;\; g_s \, h \,, \quad \quad e^{2b} \;\; \leftrightarrow \;\; f_2 \,, \quad \quad e^{2c} \;\; \leftrightarrow \;\; f_3 \,, \quad \quad k \;\; \leftrightarrow \;\; \tilde{k} \,,
\end{eqnarray}
while their 4D metric function $G_{ij}$ becomes $G \eta_{ij}$ for some function $G(r)$ due to Poincar\'e invariance, and then we have chosen to use coordinate invariance to set $G=1$.  Moreover we have the parameter relations
\begin{eqnarray}
g_s  \;\; \leftrightarrow \;\; p_0 \,, \quad \quad \bar{M} \;\; \leftrightarrow \;\; P \,, \quad \quad \bar{N} \;\; \leftrightarrow \;\; K_0 \,.
\end{eqnarray}
The ABY equations of motion and linearized solutions to order $\rho^4$ ($1/r^4$) are presented in their Appendix A.  They look formidably complicated, but simplify enormously upon imposition of four-dimensional Poincar\'e-invariance.  Additionally, many of the parameters of their ansatz (including all multiple powers of logs) can be removed using a gauge redefinition
\begin{eqnarray}
\rho \to \rho \left[ 1 + \rho^2 ( \delta_{20} + \delta_{21} \log \rho ) + \rho^4 ( \delta_{40} + \delta_{41} \log \rho + \delta_{42} \log^2 \rho + \delta_{43} \log^3 \rho \right] \,,
\end{eqnarray}
where the $\delta$s are parameters of the transformation.  Six parameters can thus be gauged away explicitly, and then a large number of other parameters are set to zero by the equations of motion.  Our solution precisely coincides with theirs with the two parameters ${\cal S}$, $\phi$ corresponding to their modes $a^{4,0}$ and $p^{4,0}$:
\begin{eqnarray}
{\cal S} \;\;  \leftrightarrow \;\; a^{4,0} \,, \quad \quad \phi \;\; \leftrightarrow \;\; p^{4,0} \,,
\end{eqnarray}
and with the other nonzero coefficients determined in terms of these as
\begin{eqnarray}
p^{(4,1)} &=& 3 \,a^{(4,0)} \,,\\
h^{(4,1)} &=& - {1\over 4} P^2 p_0 \,a^{(4,0)} \\
h^{(4,0)} &=&  {1 \over 64} (2 K_0 + 13 P^2 p_0)\, a^{(4,0)} - {1 \over 16} P^2 p_0\, p^{(4,0)} \\
K^{(4,1)} &=& -3 P^2 p_0 \,a^{(4,0)} \\
K^{(4,0)} &=& {1 \over 8} (6 K_0 + 11 P^2 p_0)\, a^{(4,0)} - {1 \over 2} P^2 p_0 \,p^{(4,0)}  \,.
\end{eqnarray}

%\newpage

%\bibliography{antibib}

\end{document}